\documentclass[suppldata]{interact} 

\usepackage{dsfont}
\usepackage{lscape,nccmath,enumitem,verbatim,bm}
\usepackage{mathrsfs} 
\usepackage{pifont}
\usepackage[scr=boondoxo]{mathalfa}
\usepackage{multirow}
\usepackage{epsfig,float}
\usepackage{tikz}
\usepackage{tabularray}
\usepackage{tabls}
\usepackage{graphicx}
\usepackage{epstopdf}
\usepackage{subfigure}

\usepackage{natbib}
\bibpunct[, ]{(}{)}{;}{a}{}{,}

\renewcommand\bibfont{\fontsize{10}{12}\selectfont}

\theoremstyle{plain}
\newtheorem{theorem}{Theorem}[]

\newtheorem{corollary}{Corollary}[]

\theoremstyle{definition}

\theoremstyle{remark}

\def\ds{\rule{0pt}{1.6ex}}

\newcommand{\bzero}{{\bf 0}}

\newcommand{\Cov}{\textrm{Cov}}
\newcommand{\Var}{{\mathrm{Var}}}

\newcommand{\bbR}{\mathbb{R}}

\begin{document}


\title{A Comprehensive Framework for Statistical Inference in Measurement System Assessment Studies}

\author{
\name{Banafsheh Lashkari\thanks{CONTACT Banafsheh Lashkari Email: blashkari@uwaterloo.ca} and Shojaeddin Chenouri}
\affil{University of Waterloo, Waterloo, ON, Canada}
}

\maketitle

\begin{abstract}
Measurement system analysis aims to quantify the variability in data attributable to the measurement system and evaluate its contribution to overall data variability. This paper conducts a rigorous theoretical investigation of the statistical methods used in such analyses, focusing on variance components and other critical parameters. While established techniques exist for single-variable cases, a systematic theoretical exploration of their properties has been largely overlooked. This study addresses this gap by examining estimators for variance components and other key parameters in measurement system assessment, analyzing their statistical properties, and providing new insights into their reliability, performance, and applicability.
\end{abstract}


\section{Introduction}

Maintaining and continuously improving quality is essential across all industries, as critical decisions regarding processes and products depend heavily on the accuracy and reliability of the measurement data~\citep{montgomery2020introduction}. However, measurement data often deviate from the true values due to a range of factors, including the limitations of the measurement devices, operator proficiency, environmental conditions, inherent item variability, and fluctuations across cycles, days, or machines~\citep{wheeler1989evaluating, taver1995manufacturing}. As such, a comprehensive understanding of these error sources is essential to ensure that decisions based on the data are sound and reliable~\citep{smith2007gauge}.\smallskip

All components involved in the process of data collection--including instruments, operators, methods, software, environmental conditions, and the broader measurement environment--constitute the measurement system~\citep{automotive1995measurement, montgomery2020introduction}. A measurement system analysis serves to quantify the variability in measurement data attributable to these components \citep{montgomery1993gaugeI}. The primary objectives of this analysis are to assess the measurement system's contribution to overall data variability, identify specific sources of variability within the measurement system, and assess its suitability for the intended application~\citep{montgomery1993gaugeII, burdick2003review, majeske2012approving}.\smallskip

The standard approach to assessing a measurement system \citep{montgomery1993gaugeI, burdick2003review} involves randomly selecting a sample of units from the process and having each operator, or more broadly, each measurement system, perform repeated measurements on these units. Here, a \emph{unit} refers to the object or item being measured. Data from this experiment are typically analyzed using a two-way model, which decomposes the measurement error into two primary components: the systematic error attributed to operator differences (reproducibility) and the error arising when the same measurement system is used multiple times over a short time period (repeatability). \smallskip

Measurement systems that are unaffected by operator bias or human factors such as skill, knowledge, or attitude are commonly referred to as measurement systems with no operator effect. These systems are typically automated, relying on machine-controlled processes like automated gauges or scanners to conduct measurements. Such systems are particularly important in industries like smart manufacturing \citep{suriano2015progressive, yang2021data, majeske2012approving}, aerospace, and the automotive sector \citep{drouot2018measurement, browne2009improved}. By assuming the operator effect is negligible or non-existent, a simplified one-way model can be employed to capture repeatability. Notable studies in measurement system assessment that utilize one-way models include those by \cite{browne2009improved, browne2010optimal, weaver2012bayesian}. \smallskip

The field of measurement system assessment has experienced significant methodological advancements, particularly for the evaluation of single characteristics. The guidelines and standards developed by~\cite{automotive2003measurement} provide a comprehensive framework to perform measurement system analyses and interpret the results. The practical considerations involved in designing and implementing measurement system studies are examined by~\cite{montgomery1993gaugeI}, while \cite{burdick2003review} offer a detailed review of methodologies for evaluating measurement system capability, with a specific focus on the ANOVA procedure. Furthermore, alternative plans to the standard measurement system assessment plan are proposed by~\cite{browne2010leveraged,stevens2010augmented,stevens2013gauge}. \smallskip

The scope of measurement system analysis methodologies has also expanded to include setups involving non-numerical scales. For instance, \cite{van2008measurement} and \cite{danila2010assessment} present frameworks for assessing measurement systems operating on binary scales, where units are classified as pass or fail.
Similarly,~\cite{culp2018analysis} propose a methodology for analyzing measurement systems with ordinal measurements, while~\cite{osthus2021gauge} develop an approach for assessing repeatability and reproducibility in systems using count data.\smallskip

Despite the field's significant methodological breadth, to the best of the authors' knowledge, there remains a lack of documented theoretical analysis of these methods and their properties in the literature. This paper addresses this gap by providing a comprehensive theoretical investigation of measurement system assessment studies for univariate data. Specifically, we examine statistical properties and considerations related to measurement system analysis parameters, focusing on the one-way random effects model under a balanced standard plan. While the one-way model is valued for its simplicity, our analysis uncovers complexities in certain aspects. Nonetheless, the insights gained from this investigation offer a solid foundation for extending these theoretical analyses to more complex scenarios, such as two-way analysis of variance, advanced models, and diverse data types, including multivariate and functional data. This paper, therefore, concentrates on the one-way random effects model, which often suffices for many practical measurement system assessment applications. \smallskip

This paper is organized as follows: Section~\ref{sec:onewaymodel} presents the univariate one-way ANOVA model, which serves as the statistical framework for evaluating measurement system variability. Section \ref{sec:parameters} outlines key parameters for assessing measurement systems and defines the foundational notation used throughout the paper. These parameters are analyzed in greater detail in subsequent sections. Section~\ref{sec:preliminary}, provides a preliminary analysis of the univariate one-way ANOVA model, laying the groundwork for a deeper understanding of the measurement system. DSection~\ref{sec:estimation} outlines various techniques for estimating the relevant parameters. A comprehensive examination of the statistical properties associated with these parameters, including estimator bias, sampling variance, hypothesis testing, convergence properties, and confidence intervals, is presented in Section~\ref{sec:statistical}. Conclusions are drawn in Section~\ref{sec:Conclusion}. Finally, the appendix contains detailed proofs that support the analysis of the paper.

\section{The univariate one-way model}
\label{sec:onewaymodel}


We consider a setting where an experiment is conducted with a randomly selected sample of $a>2$ units drawn from a population of units. Each unit in the sample is measured $r>1$ times, with the order of measurements determined randomly. See, e.g.,~\cite[Section~3]{automotive1995measurement}. It is assumed that the measurand remains constant across repeated measurements. The model used to express the measured values is, 
\begin{eqnarray}\label{eq:one-way-model}
Y_{ij} = \mu + U_i + \epsilon_{ij}, \quad\textrm{for \;} i = 1,\ldots,a, \textrm{ and }\, j=1,\ldots,r,
\end{eqnarray}
where $Y_{ij}$ is the $j$th repeated measurement for the $i$th unit, $\mu$ is a general mean parameter, $U_i$ represents the random main effect of unit $i$, and $\epsilon_{ij}$ represents the random measurement error associated with the $j$th replicate measurement taken from unit $i$. All $U_i$'s are assumed to be independent and identically distributed (i.i.d.) $\mathcal{N}(0,\sigma^2_u)$ random variables, and all $\epsilon_{ij}$'s are i.i.d. $\mathcal{N}(0,\sigma^2_\epsilon)$ random variables. Furthermore, for $i$ fixed, we treat each $U_i$ and $\epsilon_{ij}$ as being mutually independent. 

Under the one-way model of \eqref{eq:one-way-model},  while the random variables $U_i$'s and $\epsilon_{ij}$'s are uncorrelated, the measurements $Y_{ij}$'s are not uncorrelated for all $i$ and $j$ cases. Specifically, for repeated measurements $j$ and $j'$ of the same unit $i$, the covariance is given by, 
\begin{align}\label{eq:cov-in-group}
    \Cov(Y_{ij},Y_{ij'}) = \sigma^2_u \quad\textrm{for \;} j\neq j'\,,
\end{align}
whereas for measurements taken from different units, the covariance is, 
\begin{align}\label{eq:cov-bet-group}
    \Cov(Y_{ij},Y_{i'j'}) = 0 \quad\text{for \;} i\neq i', \text{ and any\;} j, j' .
\end{align}
As $Y_{ij}$'s are normally distributed random variables, the zero covariance of \eqref{eq:cov-bet-group} signifies the independence of measurements taken from different units. In contrast, the repeated measurements made on the same unit are dependent.
In view of \eqref{eq:one-way-model}, the variance of a single measurement $Y_{ij}$ is, 
\begin{align}
    \sigma^2_t = \sigma^2_u+\sigma^2_\epsilon\,.
\end{align}
The variance quantities $\sigma^2_u$ and $\sigma^2_\epsilon$ which sum to the total measurement variance $\sigma^2_t$ are referred to as variance components.

\section{Parameters for assessing measurement system}
\label{sec:parameters}
This section reviews fundamental parameters for measurement system analysis, tailored specifically to the one-way random effects model.

 The percentage R\&R ratio, as defined by AIAG \citeyearpar{automotive1995measurement}, is given by
\begin{align}\label{eq:GRR}
\%\textrm{R\&R} = \mfrac{\sigma_{\epsilon}}{\sigma_{t}} \, 100\%\,.
\end{align}
According to the guidelines set by AIAG~\citeyearpar{automotive1995measurement}, a measurement system is considered acceptable if the R\&R ratio is below $10\%$. For ratios between $10\%$ and $30\%$, the measurement system's acceptability will depend on contextual factors such as the importance of the application and the cost of implementing improvements. If the R\&R ratio exceeds $30\%$, corrective actions to enhance the measurement system are strongly recommended.

\cite{larsen2002measurement} and~\cite{burdick1997confidence} note the signal-to-noise ratio as, 
\begin{eqnarray}\label{eq:SNR}
\textrm{SNR} = \frac{\sigma_u}{\sigma_{\epsilon}}.
\end{eqnarray}
The AIAG~\citeyearpar{automotive1995measurement} and~\cite{burdick2003review}
employ a version of \eqref{eq:SNR} scaled by $\sqrt{2}$, often referred to as the gauge discrimination ratio. This ratio assesses whether the measurement system's resolution is adequate to effectively monitor the feature of interest. According to AIAG guidelines, an acceptable value for the gauge discrimination ratio is $5$ or greater, while a value below $2$ indicates that the measurement system is unsuitable for process monitoring. Steiner and MacKay~\citeyearpar{steiner2005statistical} propose more moderate criteria, suggesting that an SNR value exceeding $3$, as defined by \eqref{eq:SNR}, indicates a valid measurement system. Conversely, a value below $2$ is considered unacceptable.


An alternative ratio to compare the quality of a measurement system is the intra-class correlation coefficient (ICC), defined as~\citep{mader1999economic,majeske2002evaluating},  
\begin{eqnarray}\label{eq:ICC}
\textrm{ICC}=\frac{\sigma^2_u}{\sigma^2_t}  = \frac{\sigma^2_u}{\sigma^2_u + \sigma^2_{\epsilon}}.
\end{eqnarray}
It reflects the correlation in repeated measurements on the same subject~\citep{majeske2002evaluating}, and~\cite{donner1987sample} regard the intra-class correlation coefficient as a measure of reliability. When the variability of the measurement system is negligible relative to that of the units, the ICC approaches one, indicating high reliability. Conversely, when the measurement system's variability is substantial compared to the variability of units, the ICC decreases, signifying lower reliability.


The AIAG~\citeyearpar{automotive1995measurement} uses the {precision-to-tolerance ratio} (PTR) as an additional approval criterion. PTR is defined as the ratio of the width of the probability distribution of repeated measurements on the same unit to the tolerance width of the process. If ${UL}$ and ${LL}$ represent the upper and lower specification limits of the process, the PTR is expressed as 
\begin{eqnarray}\label{eq:PTR}
\textrm{PTR} = \frac{\kappa\,\sigma_{\epsilon}}{UL-LL}\, ,
\end{eqnarray}
where $\kappa$ is a scaling factor commonly set to $5.15$ or $6$, corresponding to capturing $99\%$ and $99.73\%$ of a normal distribution with variance $\sigma^2_{\epsilon}$, respectively.

The literature suggests varying thresholds for acceptable PTR values, typically ranging from $0.1$ to $0.3$. For instance, \cite{montgomery1993gaugeI} recommend a PTR value of $0.1$ or less for a measurement system to be considered adequate.~\cite{mader1999economic} propose a maximum PTR value of $0.2$, while \cite{barrentine2003concepts} suggests that a PTR value not exceed $0.3$. However,~\cite{montgomery1993gaugeI} caution against over-relying on PTR. They argue that a measurement system's ability to detect meaningful variability in units is more important. 
Moreover, they emphasize that metrics such as \%R\&R and SNR provide more comprehensive insights into the measurement system's performance than PTR alone.


Throughout this paper, the quality of the measurement system is assessed using the ratio of the two variance components,
\begin{align}\label{eq:rho}
\rho = \frac{\sigma^2_u}{\sigma^2_\epsilon}.
\end{align}
The parameters introduced earlier to evaluate the measurement system—namely, the percentage R\&R ratio, $\text{SNR}$, and the intra-class correlation coefficient (ICC)—are related to $\rho$ through
$ \%\textrm{R\&R}=\left(1+\rho\right)^{-1/2} \times 100\% $, $\text{SNR} = \rho^{1/2}$, and $\textrm{ICC} =  1/\left(1+\rho^{-1}\right)$.\smallskip

A smaller value of $\rho$ suggests that measurement error is the dominant source of variation, whereas larger values of $\rho$ indicate that measurement error contributes less significantly to the overall variability in the observed data. It is important to note that the recommended range for $\rho$ can vary depending on the specific context. For instance, as pointed out by~\cite{steiner2005statistical}, improving the measurement system is advisable if $\rho$ falls below approximately $4$.


\section{Preliminary analysis}
\label{sec:preliminary}

In the context of the one-way model \eqref{eq:one-way-model}, the two sources of variation are the units~($U$) and the measurement errors~($\epsilon$). These sources represent the underlying factors or components that contribute to the variation observed in the measurement data~($Y$).

Two sums of squares that are the basis for the analysis of the variance components of the one-way model \eqref{eq:one-way-model} are,
\begin{align}\label{eq:SS_S}
    SS_u &= r\sum_{i=1}^a (\overline{Y}_{i\cdot} - \overline{Y}_{\cdot\cdot})^2, \\   
\label{eq:SS_E}
 SS_\epsilon &=\sum_{i=1}^a\sum_{j=1}^r (Y_{ij} - \overline{Y}_{i\cdot})^2\, ,
\end{align}
where $\overline{Y}_{i\cdot} = \frac{1}{r}\sum_{j=1}^r Y_{ij}$ is the  mean of the measurements of the $i$th unit, and $\overline{Y}_{\cdot\cdot} = \frac{1}{ar}\sum_{i=1}^a\sum_{j=1}^r Y_{ij}$ is the overall mean of measured data. There are two means of squares,
\begin{align}\label{eq:MS_S}
    {MS}_u &= \mfrac{r}{a-1}\sum_{i=1}^a (\overline{Y}_{i\cdot} - \overline{Y}_{\cdot\cdot})^2, \\
\label{eq:MS_E}
     MS_\epsilon &= \mfrac{1}{a(r-1)}\sum_{i=1}^a\sum_{j=1}^r (Y_{ij} - \overline{Y}_{i\cdot})^2\,.
\end{align}
The total corrected sum of squares is defined as
\begin{align}
    SS_t=\sum_{i=1}^a\sum_{j=1}^r (Y_{ij} - \overline{Y}_{\cdot\cdot})^2.
\end{align}
This total sum of squares can be partitioned into,
\begin{align}
    SS_t=SS_u+SS_\epsilon\,.
\end{align}
This classification based on the source of variation is known as the {analysis of variance} (ANOVA). The ANOVA classification associated with the one-way model of \eqref{eq:one-way-model} is demonstrated in Table~\ref{table:ANOVA-oneway}.

\begin{table}[!htb]
    \centering
    \caption{The analysis of variance classification for one-way model~\eqref{eq:one-way-model}.\vspace{2mm}}
    \scalebox{0.8}{
    \begin{tblr}{
    colspec ={c c c c},
    hline{1} = {},
    vline{1-5} ={},
    cell{1}{1-4} = {blue!10},
    }
 Source &  Degree of freedom & {Sum of squares}  & Mean of squares \\ [0.2ex] 
 \hline\hline  
  $U$ &  $\quad \mathrm{df}_u=a-1 $ & $SS_u = r\sum_{i=1}^a (\overline{Y}_{i\cdot} - \overline{Y}_{\cdot\cdot})^2$ & $MS_u = \mfrac{SS_u}{a-1}$  \\ 
  $\epsilon$ & $\quad \mathrm{df}_\epsilon=a(r-1)$ & $SS_{\epsilon}=\sum_{i=1}^a\sum_{j=1}^r (Y_{ij} - \overline{Y}_{i\cdot})^2$  & $MS_{\epsilon} = \mfrac{SS_{\epsilon}}{a(r-1)}$ \\ \hline
  Total & $ar-1$ & $SS_t=\sum_{i=1}^a\sum_{j=1}^r (Y_{ij} - \overline{Y}_{\cdot\cdot})^2$ &{} \\ [0.2ex] 
    \hline
\end{tblr}
\label{table:ANOVA-oneway}
}
\end{table}


With the normal distribution of $U_i$'s and $\epsilon_{ij}$'s and their mutual independence, $SS_u$ and $SS_\epsilon$ are two independent statistics where, 
\begin{align}\label{eq:SS}
    \frac{SS_u}{\mathrm{E}\left[MS_u\right]}\sim \chi^2_{a-1}\,,\quad
\mbox{and }\quad  \frac{SS_\epsilon}{\mathrm{E}\left[MS_\epsilon\right]}\sim \chi^2_{a(r-1)}\,.
\end{align}
The expected mean of squares terms are, 
\begin{align}
\mathrm{E}\left[MS_u\right] =\sigma^2_\epsilon + r \sigma^2_u \quad
\mbox{and }\quad \mathrm{E}\left[MS_{\epsilon}\right] =\sigma^2_\epsilon \,.
\end{align}
The distributions of $SS_u$ and $SS_\epsilon$ given in~\eqref{eq:SS} lead to the result that
\begin{align}\label{eq:F}
    F=\frac{{MS_u}/{\mathrm{E}\left[MS_u\right]}}{{MS_\epsilon}/{\mathrm{E}\left[MS_{\epsilon}\right]}}={(1+r\rho)^{-1}}\frac{MS_u}{MS_\epsilon}
\end{align}
has a central $F$-distribution with $a-1$ and $a(r-1)$ degrees of freedom. 

To initiate a measurement system study,~\cite{montgomery1993gaugeI} recommend starting with an initial analysis to identify the factors contributing to the variation in the measured data. This can be achieved by conducting a hypothesis test for $\sigma^2_u,$. Once the significance of this variance component is statistically confirmed, the variance components can then be estimated.


\section{Point estimation}
\label{sec:estimation}
The estimation of variance components for the one-way model has been extensively studied, with a rich body of literature available on the topic. For a thorough review of methods in this area, readers are directed to references such as~\cite{klotz1969mean,sahai1979bibliography,khuri1985variance,robinson1987estimation,searle1992variance}. This section provides a concise overview of several established techniques for estimating variance components within the context of the one-way model defined by Equation~\eqref{eq:one-way-model}. Specifically, we explore the ANOVA-based estimators, the uniformly minimum variance unbiased estimator (UMVUE), and the maximum likelihood estimator. Additionally, we employ the commonly used plug-in estimator to derive estimates for ratios of variance components, which are integral to measurement system assessment studies.


\subsection{ANOVA estimation}
A common approach for estimating variance components is the ANOVA method, which is conceptually similar to the method of moments. In this procedure, each sum (or mean) of squares has an expected value that is a linear function of the variance components. The ANOVA estimation method solves these equations for $\sigma^2_u$ and $\sigma^2_\epsilon$, substituting the expected mean squares with their observed values. Consequently, the ANOVA estimators for $\sigma^2_u$ and $\sigma^2_\epsilon$ are expressed as,
\begin{align}\label{eq:ANOVA-est}
\widehat{\sigma^2_u} = \mfrac{1}{r}(MS_u-MS_\epsilon) 
\,\qquad\mbox{and}\qquad 
\widehat{\sigma^2_\epsilon\,} = MS_\epsilon\,.
\end{align}
These estimators are unbiased. By substituting the ANOVA estimators from \eqref{eq:ANOVA-est} into \eqref{eq:rho}, the plug-in estimator of $\rho$ is given by,
\begin{align}\label{eq:ANOVA-rho}
    \widehat{\rho}_{\ds{\tiny\mbox{ANOVA}}}=\mfrac{1}{r}\left(\mfrac{MS_u}{MS_\epsilon}-1\right).
\end{align}

One drawback of the ANOVA estimator is its potential to yield a negative estimate for the variance component $\sigma^2_u$. Specifically, the estimator $\widehat{\sigma^2_u}$, as defined in \eqref{eq:ANOVA-est}, will produce a negative value whenever $MS_u<MS_\epsilon\,$. The likelihood of this occurrence depends on the characteristics of the data. For the one-way model in \eqref{eq:one-way-model}, the probability of this event can be expressed as,
\begin{align}\label{eq:chap2-prob-neg}
    \Pr\left({MS_u}<{MS_\epsilon}\right)=\Pr\left(F<{(1+r\rho)^{-1}}\right),
\end{align}
where the random variable $F$ follows an $F$-distribution with $a-1$ and $a(r-1)$ degrees of freedom. 

The probability defined in~\eqref{eq:chap2-prob-neg} depends on the sample size $a$, the number of measurement replications $r$, and the ratio of variance components $\rho$. 
A lower probability of $MS_u<MS_\epsilon$ is preferred, as it indicates a higher likelihood that the ANOVA estimates of variance components align with the parameter space requirements, specifically \mbox{$\sigma^2_u\in\bbR^{0+}$}, and $\sigma^2_\epsilon \in \bbR^+$, where $\mathbb{R}^{0+}$ and $\mathbb{R}^+$ denote the sets of non-negative real numbers and positive real numbers, respectively.

\begin{figure}[!htbp]
\centering
\subfigure[$a=5$]{%
\resizebox*{6.5cm}{!}{\includegraphics{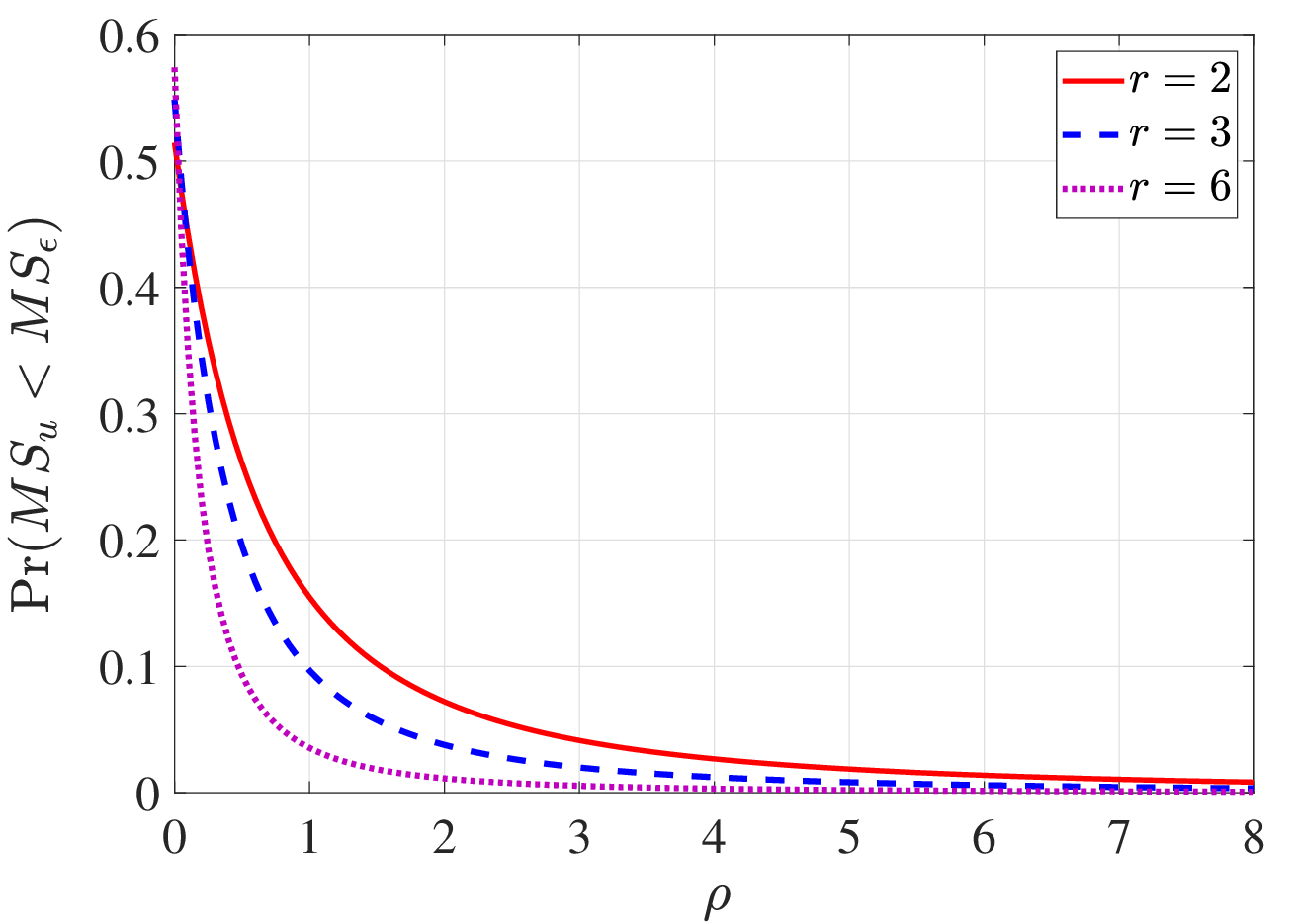}}}
\hspace{10pt}
\subfigure[$a=10$]{%
\resizebox*{6.5cm}{!}{\includegraphics{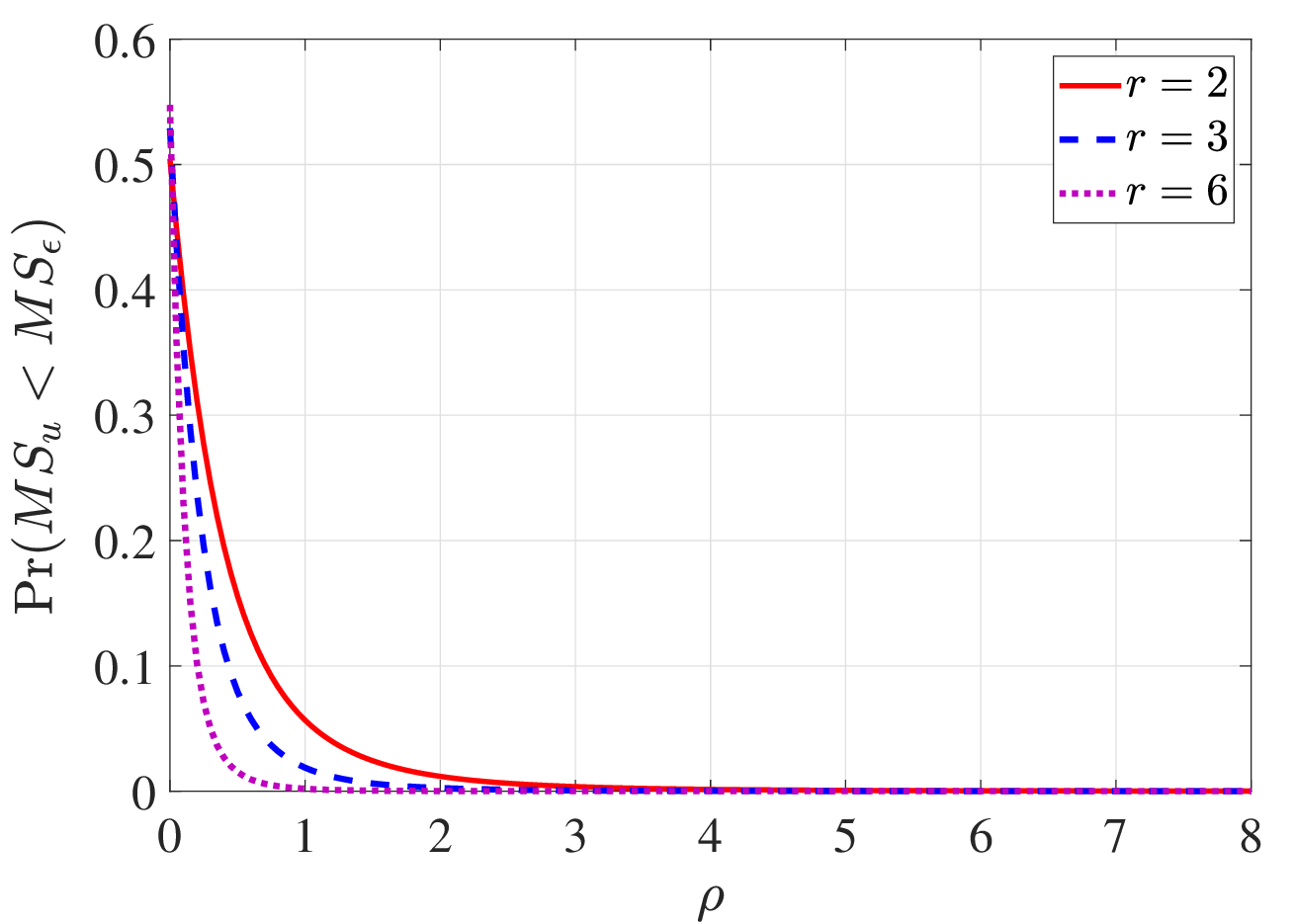}}}
\\
\subfigure[$a=20$]{%
\resizebox*{6.5cm}{!}{\includegraphics{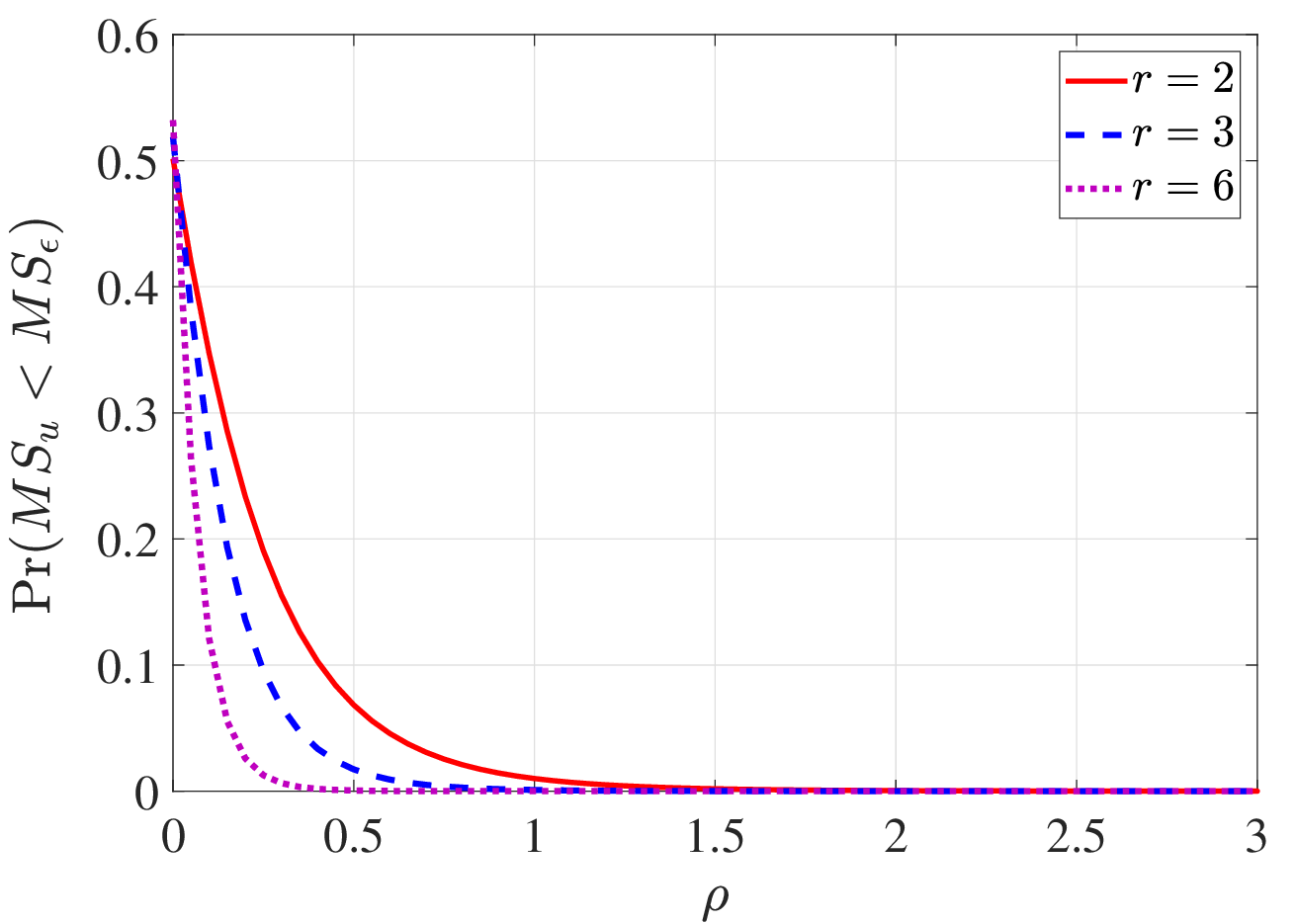}}}
\hspace{10pt}
\subfigure[$a=30$]{%
\resizebox*{6.5cm}{!}{\includegraphics{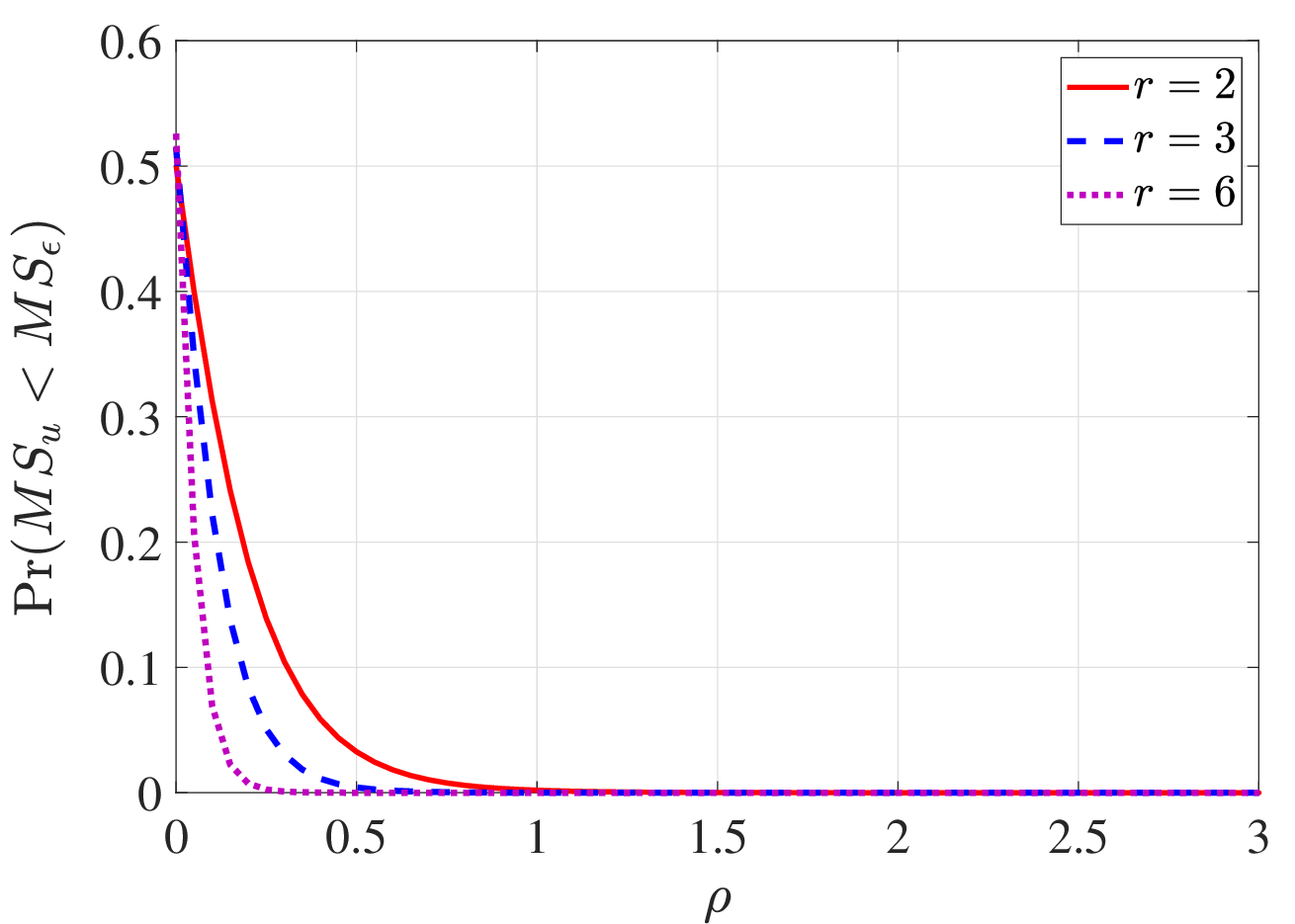}}}
\caption{The probability of $MS_u<MS_\epsilon$  as a function of $\rho$ for $a=5,10,20,30$ and $r=2,3,6$, when the units random effects $U_i$'s and measurement errors $\epsilon_{ij}$'s are normally distributed.} \label{Fig:negative-est} 
\end{figure}

To better understand the likelihood of this scenario, Figure~\ref{Fig:negative-est} illustrates the relationship between the probability of $MS_u<MS_\epsilon\,$ and $\rho$ across four sample sizes \mbox{($a=5,10,20,30$)} and three values of replications ($r=2,3,6$), a total of $12$ plans. As shown in Figure~\ref{Fig:negative-est}, increasing the number of measurements, i.e., $N=ar$, whether by increasing the sample size $a$ or the number of replications $r$, reduces the probability of \mbox{$MS_u<MS_\epsilon\,$}. Generally, larger values of $\rho$ are more advantageous as they correspond to a lower likelihood of \mbox{$MS_u<MS_\epsilon\,$}~\citep{majeske2002evaluating}. 

A negative estimate of variance can be interpreted in different ways~\citep{thompson1962problem}. It may indicate that the variance component is not significantly different from zero, suggesting the possibility of simplifying the model. This interpretation can be statistically validated through a hypothesis test for the null hypothesis $H_{_0}\,:\,\sigma^2_u=0\,$. 

Insufficient data is another potential cause of negative variance estimates, highlighting the need to incorporate more data. In the context of measurement system assessment, baseline (or historical) data are often readily available and can be leveraged to enhance analysis. The use of baseline data is recommended in the literature~\cite{steiner2005statistical}, and its advantages are quantified by~\cite{stevens2013gauge}.


\subsection{Non-negative ANOVA estimation}
When considering $\sigma^2_u=0\,$, the ANOVA estimator of $\sigma^2_\epsilon$ is calculated as follows, 
\begin{align}
    \widehat{\sigma^2_\epsilon\,} = 
    \frac{SS_t}{ar-1}\,.
\end{align}
This expression for $\widehat{\sigma^2_\epsilon\,}$ can also be written as $(MS_u \mathrm{df}_u +MS_\epsilon \mathrm{df}_\epsilon)/\left(\mathrm{df}_u + \mathrm{df}_\epsilon\right)$, which represents a pooled variance estimator combining $MS_u$ and $MS_\epsilon\,$. 

In cases where $MS_u<MS_\epsilon\,$, it can be demonstrated that ${SS_t}/({ar-1})<MS_\epsilon\,$. Conversely, if $MS_\epsilon<MS_u\,$, then \mbox{$MS_\epsilon<{SS_t}/({ar-1})$}.
Utilizing these relationships, the non-negative ANOVA estimators of variance components are expressed as follows,
\begin{align}\label{eq:N-ANOVA}
    \widehat{\sigma^2_u} = \max\left(0,\mfrac{1}{r}\left(MS_u-MS_\epsilon\right)\right)\,\quad\mbox{and}\quad 
    \widehat{\sigma^2_\epsilon\,} = \min\left(\mfrac{SS_t}{ar-1}\,,MS_\epsilon\right).
\end{align}

These non-negative ANOVA estimators of variance components are no longer unbiased. However, they exhibit smaller mean square errors compared to their traditional ANOVA counterparts, as noted by~\cite{lee1992mean}.

The plug-in estimator of $\rho$ using the non-negative estimators of variance components is expressed as, 
\begin{align}\label{eq:N-ANOVA_rho}
    \widehat{\rho}_{\ds{\tiny\mbox{NANOVA}}} = \max\left(0,\mfrac{1}{r}\left(\mfrac{MS_u}{MS_\epsilon}-1\right)\right).
\end{align}

\subsection{Uniformly minimum variance unbiased estimators}
The UMVUEs of variance components have been developed for the one-way model. Let $\bm\theta = (\mu,\, \sigma^2_u,\, \sigma^2_\epsilon )^\top$ represent the vector of unknown parameters in the model \eqref{eq:one-way-model}. Suppose the observed data of the measurements on unit $i$ are arranged as $\bm{y}_i=(y_{i1},\ldots,y_{ir})^\top$. The likelihood function of $\bm\theta$, as set up in~\cite[Chapter~3]{searle1992variance}, is expressed by
\begin{align}\label{eq:L-one-way}
\resizebox{0.92\textwidth}{!}{
    $\mathcal{L}(\bm\theta) = c(\sigma^2_u\,,\sigma^2_\epsilon) \exp\left\lbrace\mfrac{-1}{2\sigma^2_\epsilon}{\sum\limits_{i=1}^a\sum\limits_{j=1}^r (y_{ij} - \overline{y}_{i\cdot})^2}-\mfrac{1}{2(\sigma^2_\epsilon + r \sigma^2_u)}{ \sum\limits_{i=1}^a\sum\limits_{j=1}^r (\overline{y}_{i\cdot} - \overline{y}_{\cdot\cdot})^2}\!-\mfrac{ar(\mu - \overline{y}_{\cdot\cdot} )^2}{2(\sigma^2_\epsilon + r \sigma^2_u)}\right\rbrace$}
\end{align}
where 
$ c(\sigma^2_u\,,\sigma^2_\epsilon) = (2\pi)^{-\frac{1}{2}ar}(\sigma^2_\epsilon)^{-\frac{1}{2}a(r-1)}(\sigma^2_\epsilon + r\sigma^2_u)^{-\frac{a}{2}}\,,\nonumber
$
and $\overline{y}_{i\cdot}= \frac{1}{r}\sum_{j=1}^r y_{ij}$ and $\overline{y}_{\cdot\cdot}= \frac{1}{ar}\sum_{i=1}^a\sum_{j=1}^r y_{ij}\,$. \smallskip

From the structure of the likelihood function, it can be deduced that the joint distribution of $Y_{ij}$'s belongs to the exponential family and the vector $(\overline{Y}_{\cdot\cdot}, MS_u, MS_\epsilon )^\top$ serves as a jointly sufficient and complete statistic for the parameter vector $\bm\theta$. By utilizing an unbiased estimator of $\bm\theta$, and applying the Lehmann-Scheff\'e theorem~(1950,\,1955), this estimator becomes the UMVUE for $\bm\theta$, depending solely on $(\overline{Y}_{\cdot\cdot}, MS_u, MS_\epsilon)^\top$. It is noteworthy that the ANOVA estimators of variance components, as outlined in \eqref{eq:ANOVA-est} are unbiased since  $\mathrm{E}[\widehat{\sigma^2_{u}} ]=\sigma^2_u$ and $\mathrm{E}[\widehat{\sigma^2_\epsilon\,}]=\sigma^2_\epsilon$. Consequently, the UMVUEs for variance components $\sigma^2_u$ and $\sigma^2_\epsilon$ coincide with their respective ANOVA estimators~\citep{graybill1956note}.

\subsection{Maximum likelihood estimation}
The maximum likelihood estimation method is underpinned by a strong theoretical foundation. Taking logarithm of $\mathcal{L(\bm\theta)}$, the log-likelihood function of $\bm\theta$ is, 
\begin{equation}
\begin{split}\label{eq:log-likelihood}
    \mathscr{l}(\bm\theta) = &-\mfrac{a(r-1)}{2}\ln(\sigma^2_\epsilon) - \mfrac{a}{2}\ln(\sigma^2_\epsilon + r \sigma^2_u)-\frac{1}{2\sigma^2_\epsilon}{\sum\limits_{i=1}^a\sum\limits_{j=1}^r (y_{ij} - \overline{y}_{i\cdot})^2}\\
    &-\frac{1}{2(\sigma^2_\epsilon + r \sigma^2_u)}\sum\limits_{i=1}^a\sum\limits_{j=1}^r (\overline{y}_{i\cdot} - \overline{y}_{\cdot\cdot})^2 -\frac{ar(\overline{y}_{\cdot\cdot} - \mu)^2}{2(\sigma^2_\epsilon + r \sigma^2_u)} -\mfrac{ar}{2} \ln(2\pi) \,.
\end{split}
\end{equation}
The maximum likelihood estimation involves maximizing the likelihood function over the parameter space. 
In \eqref{eq:log-likelihood}, the parameter $\mu$ appears in the quadratic term $(\overline{y}_{\cdot\cdot} - \mu)^2$, and its parameter space is $\bbR$. The log-likelihood function $\mathscr{l}(\bm\theta)$ attains its maximum when ${\mu}=\overline{y}_{\cdot\cdot}\,$.

By calculating the partial derivatives of $\mathscr{l}(\bm\theta)$ with respect to $\sigma^2_u$ and $\sigma^2_\epsilon$, and solving the resulting score equations, the maximum likelihood estimators for the variance components are obtained as,
\begin{align}\label{eq:MLE-est}
\widehat{\sigma^2_{u}} = \mfrac{1}{r}\big(\beta^{-1}
MS_u-MS_\epsilon\big)\qquad \mbox{and} \qquad
\widehat{\sigma^2_\epsilon\,} = MS_\epsilon\,,
\end{align}
where the scaling factor $\beta$ is defined as $\beta = \mfrac{a}{a-1}\,$.

The estimators in  (\ref{eq:MLE-est}) are considered as the maximum likelihood estimators under the condition that both $\widehat{\sigma^2_u}$ and $\widehat{\sigma^2_\epsilon\,}$ lie within the parameter space of $\sigma^2_u$ and $\sigma^2_\epsilon$. The estimator $\widehat{\sigma^2_\epsilon\,} = MS_\epsilon$ is inherently positive, and therefore it satisfies the parameter space requirements for~${\sigma}^2_\epsilon$. However, the estimator $\widehat{\sigma^2_u}$, as defined in (\ref{eq:MLE-est}) remains non-negative only if the condition ${MS_u} \ge \beta{MS_\epsilon}$ holds. 

When ${MS_u} < \beta {MS_\epsilon}$, the log-likelihood function $\mathscr{l}(\bm\theta)$ achieves its maximum at the boundary of the parameter space, leading to the  following estimators,
\begin{align}\label{eq:MLE-est-0}
\widehat{\sigma^2_{u}} = 0\, \qquad \mbox{and} \qquad
\widehat{\sigma^2_\epsilon\,} = \frac{SS_t}{ar}\,\cdot
\end{align}
n this case, the estimators provided in (\ref{eq:MLE-est-0}) serve as the maximum likelihood estimators for ${\sigma}^2_{u}$ and ${\sigma}^2_{\epsilon}$, respectively~\cite[Chapter~3]{searle1992variance}. 
By combining the relations~\eqref{eq:MLE-est} and \eqref{eq:MLE-est-0}, the maximum likelihood estimators for ${\sigma}^2_{u}$ and ${\sigma}^2_{\epsilon}$ can be expressed as follows,
\begin{align}\label{eq:uni-ML}
    \widehat{\sigma^2_u} = \max\left(0\,,\mfrac{1}{r}\left(\beta^{-1}MS_u-MS_\epsilon\right)\right)\quad\mbox{and}\quad 
    \widehat{\sigma^2_\epsilon\,} = \min\left(\mfrac{SS_t}{ar},MS_\epsilon\right).
\end{align}
Substituting these maximum likelihood estimators for the variance components into the equation for $\rho$, the maximum likelihood estimator for $\rho$ is given by,
\begin{align}\label{eq:MLE_rho}
    \widehat{\rho}_{_{\tiny\mbox{MLE}}} = \max\left(0\,,\mfrac{1}{r}\left(\beta^{-1}\mfrac{MS_u}{MS_\epsilon}-1\right)\right).
\end{align}

\subsection{Other estimation methods}
Other approaches are available for estimating the variance components. Two notable methods include restricted maximum likelihood \citep{thompson1962problem,patterson1971recovery} and the Bayesian framework. 

The restricted maximum likelihood method is an extension of the maximum likelihood estimation technique. It aims to estimate model parameters by maximizing the likelihood of the observed data while incorporating constraints on these parameters. This method is particularly advantageous as it provides less biased results compared to standard maximum likelihood estimation. Under the assumption of normality for the random effects and balanced study designs, restricted maximum likelihood estimators are identical to the non-negative ANOVA estimators~\citep{corbeil1976comparison}.\smallskip

Within the Bayesian framework, variance components are treated as random variables with associated prior distributions. Bayesian estimation uses Bayes' theorem to derive a conditional probability distribution, known as the posterior distribution, which integrates prior knowledge with observed data. This posterior distribution serves as the foundation for estimating the parameters of interest, often requiring numerical methods to obtain solutions~\cite[Chapter 3]{searle1992variance}. 

Prominent examples of Bayesian estimation include the works of~\cite{tiao1965bayesian,klotz1969mean,portnoy1971formal}, which investigate Bayes estimators of variance components in the context of the one-way model. Additionally, studies by~\cite{rajagopalan1983bayesian,fong2010bayesian}, extend these methods to more general linear models. The versatility of the Bayesian approach is further demonstrated by~\cite{weaver2012bayesian}, who illustrate its application across a wide range of measurement system analyses.

A summary of the estimator results discussed in this section is provided in Table~\ref{table:Estimator-results}.

\begin{table}[htb!]
\centering
\caption{Estimators of variance components and the parameter $\rho$ in a one-way random effect model with balanced data.    \vspace{2mm}}
\scalebox{0.8}{
\begin{tblr}{
    colspec={|p{4.8cm}|p{6.0cm}|p{6.0cm}|},
    hline{1,4,6,8}={},
    vline{1,2,3,4}={},
    cell{1}{1-3} = {blue!10},
    cell{1}{1-3} = {r=1}{valign = m},
    cell{2}{1,3}={r=2}{valign=m},
    cell{4}{1,3}={r=2}{valign=m},
    cell{6}{1,3}={r=2}{valign=m},
    }
 Method &   {Estimators of variance components}   &  {Estimator of parameter $\rho$}\\ \hline \hline [0.2ex] 
  {ANOVA  and UMVUE } &   $\widehat{\sigma^2_u} = \mfrac{1}{r}(MS_u-MS_\epsilon) $ & $ \widehat{\rho}_{\ds{\tiny\mbox{ANOVA}}}=\mfrac{1}{r}\left(\mfrac{MS_u}{MS_\epsilon}-1\right)$ \\
  & $\widehat{\sigma^2_\epsilon\,} = MS_\epsilon$  & \\  
  Non-negative ANOVA  and restricted maximum likelihood &  $\widehat{\sigma^2_u} = \max\left(0,\mfrac{1}{r}(MS_u-MS_\epsilon)\right)$  & $\widehat{\rho}_{\ds{\tiny\mbox{NANOVA}}} = \max\left(0,\mfrac{1}{r}\left(\mfrac{MS_u}{MS_\epsilon}-1\right)\right)$\\
  & $\widehat{\sigma^2_\epsilon\,} = \min\left(\mfrac{SS_t}{ar-1}\,,MS_\epsilon\right)$ & \\
  Maximum likelihood &  $\widehat{\sigma^2_u} = \max\left(0\,,\mfrac{1}{r}\left(\beta^{-1}MS_u-MS_\epsilon\right)\right)$ & $\widehat{\rho}_{_{\tiny\mbox{MLE}}} = \max\left(0\,,\mfrac{1}{r}\left(\beta^{-1}\mfrac{MS_u}{MS_\epsilon}-1\right)\right)$\\
  & $\widehat{\sigma^2_\epsilon\,} = \min\left(\mfrac{SS_t}{ar}\,,MS_\epsilon\right)$ &  
\end{tblr}
}
\label{table:Estimator-results}
\end{table}


\section{Theoretical properties}\label{sec:statistical}
The estimation of measurement system study parameters necessitates a thorough understanding of their statistical properties. In this section, we conduct an in-depth analysis of the statistical properties of variance components and their ratio within the context of a measurement system assessment study. This exploration offers valuable insights into the fundamental characteristics and overall performance of the measurement system.

\subsection{The expected values and bias}
In our theoretical analysis, we begin by investigating the expected value and bias associated with the estimation of variance components and their ratio. Our focus centers on conducting a detailed examination of these properties using estimators derived from the ANOVA, non-negative ANOVA, and maximum likelihood methods.

\subsubsection*{A. ANOVA estimations}
The ANOVA estimators of variance components, as provided in \eqref{eq:ANOVA-est}, are unbiased since \mbox{$\mathrm{E}[MS_u ]=r\sigma^2_u + \sigma^2_\epsilon$} and $\mathrm{E}[MS_{\epsilon} ]=\sigma^2_\epsilon$.  However, this unbiasedness does not extend directly to the estimator of $\rho$ as defined by~\eqref{eq:ANOVA-rho}. The expected value of $\widehat{\rho}_{\ds{\tiny\mbox{ANOVA}}}$ according to~\eqref{eq:ANOVA-rho} is expressed as follows,
\begin{align}\label{eq:ch2-ANOVA-Erho}
    \mathrm{E}\left[\widehat{\rho}_{\ds{\tiny\mbox{ANOVA}}}\right]
    &=\mfrac{1}{r}\mathrm{E}\left[\mfrac{MS_u}{MS_\epsilon}\right]-\mfrac{1}{r}\nonumber\\
    &=\mfrac{1}{r}(1+r\rho)\mathrm{E}[F]-\mfrac{1}{r},
\end{align}
where $F$ is a random variable that has a central $F$-distribution with $\mathrm{df}_u=a-1$ and $\mathrm{df}_\epsilon=a(r-1)$ degrees of freedom. Note that the expectation of $F$ is $\mathrm{df}_\epsilon/(\mathrm{df}_\epsilon-2)$ where $\mathrm{df}_\epsilon >2$. Subsequently, the expected value of $\widehat{\rho}_{\ds{\tiny\mbox{ANOVA}}}$ can be expressed as,
\begin{align}\label{eq:mean-rho}
    \mathrm{E}\left[\widehat{\rho}_{\ds{\tiny\mbox{ANOVA}}}\right]
    &=\frac{1}{\mathrm{df}_\epsilon-2}\left(\rho\, \mathrm{df}_\epsilon +\mfrac{2}{r}\right).
\end{align}
The difference between the expected value of an estimator and the true value of the parameter it estimates is defined as the bias of the estimator. For $\widehat{\rho}_{\ds{\tiny\mbox{ANOVA}}}$, the bias is given by,
\begin{align}
    \mathrm{Bias}\left[\widehat{\rho}_{\ds{\tiny\mbox{ANOVA}}}\right]=\mathrm{E}\left[\widehat{\rho}_{\ds{\tiny\mbox{ANOVA}}}\right] - \rho =\frac{2(1+r\rho)}{r(\mathrm{df}_\epsilon-2)}
\end{align}
This equation reveals that $\widehat{\rho}_{{\tiny\mbox{ANOVA}}}$ consistently exhibits a positive bias. However, the bias diminishes as the sample size or the number of replications increases, eventually approaching zero.

\begin{figure}[!h]
\centering
\subfigure[$a=5$]{%
\resizebox*{50mm}{!}{\includegraphics{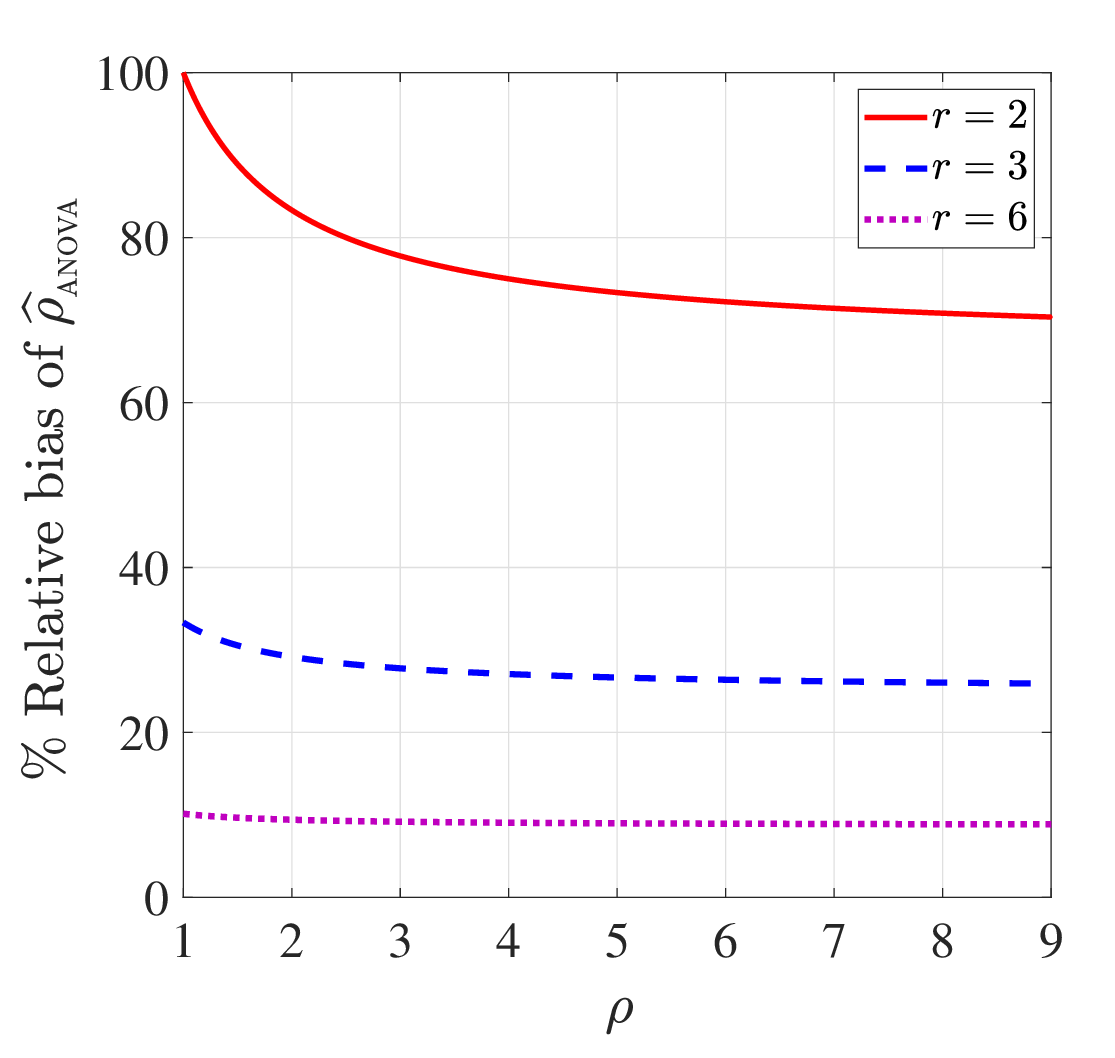}}}
\hspace{5pt}
\subfigure[$a=10$]{%
\resizebox*{50mm}{!}{\includegraphics{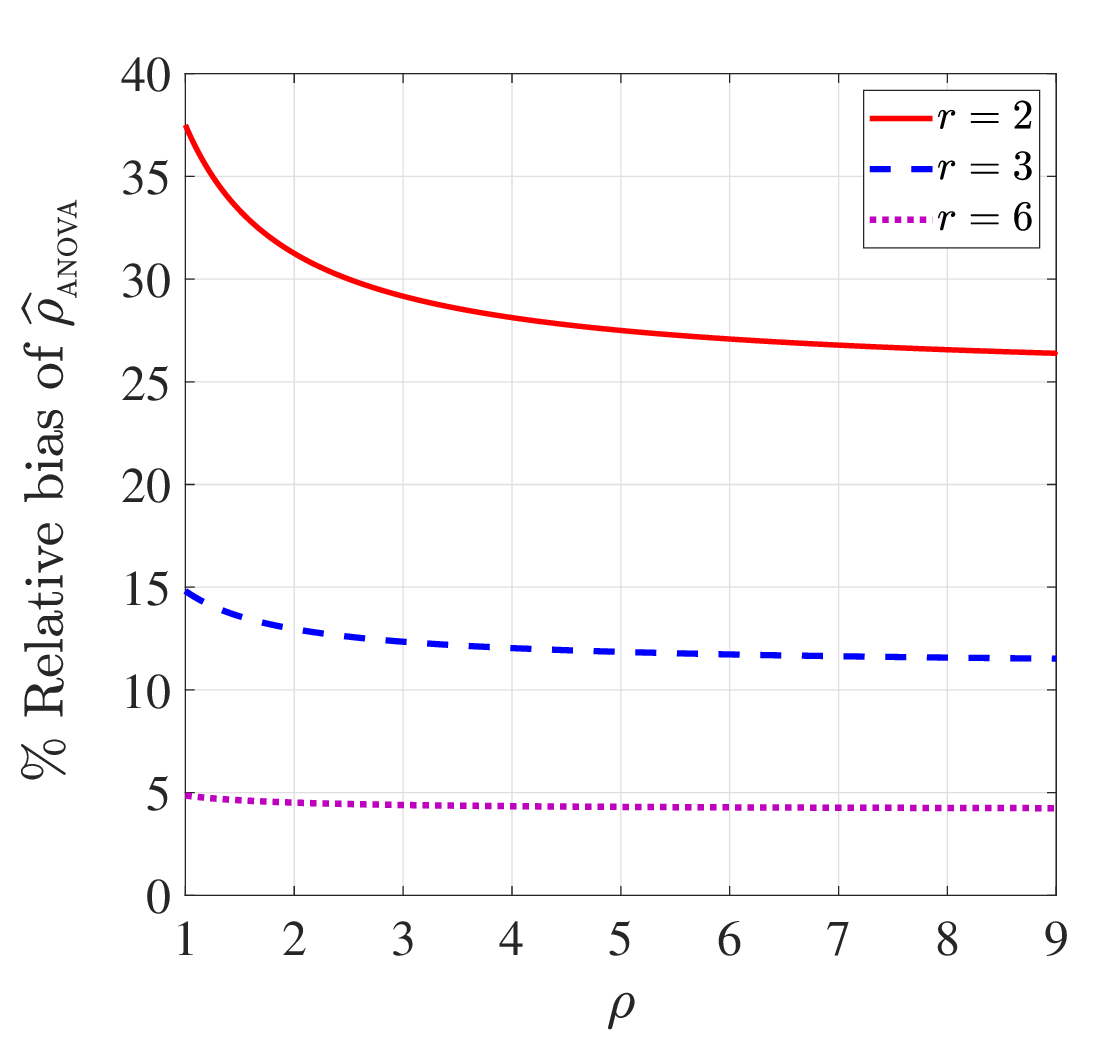}}}
\\
\subfigure[$a=20$]{%
\resizebox*{50mm}{!}{\includegraphics{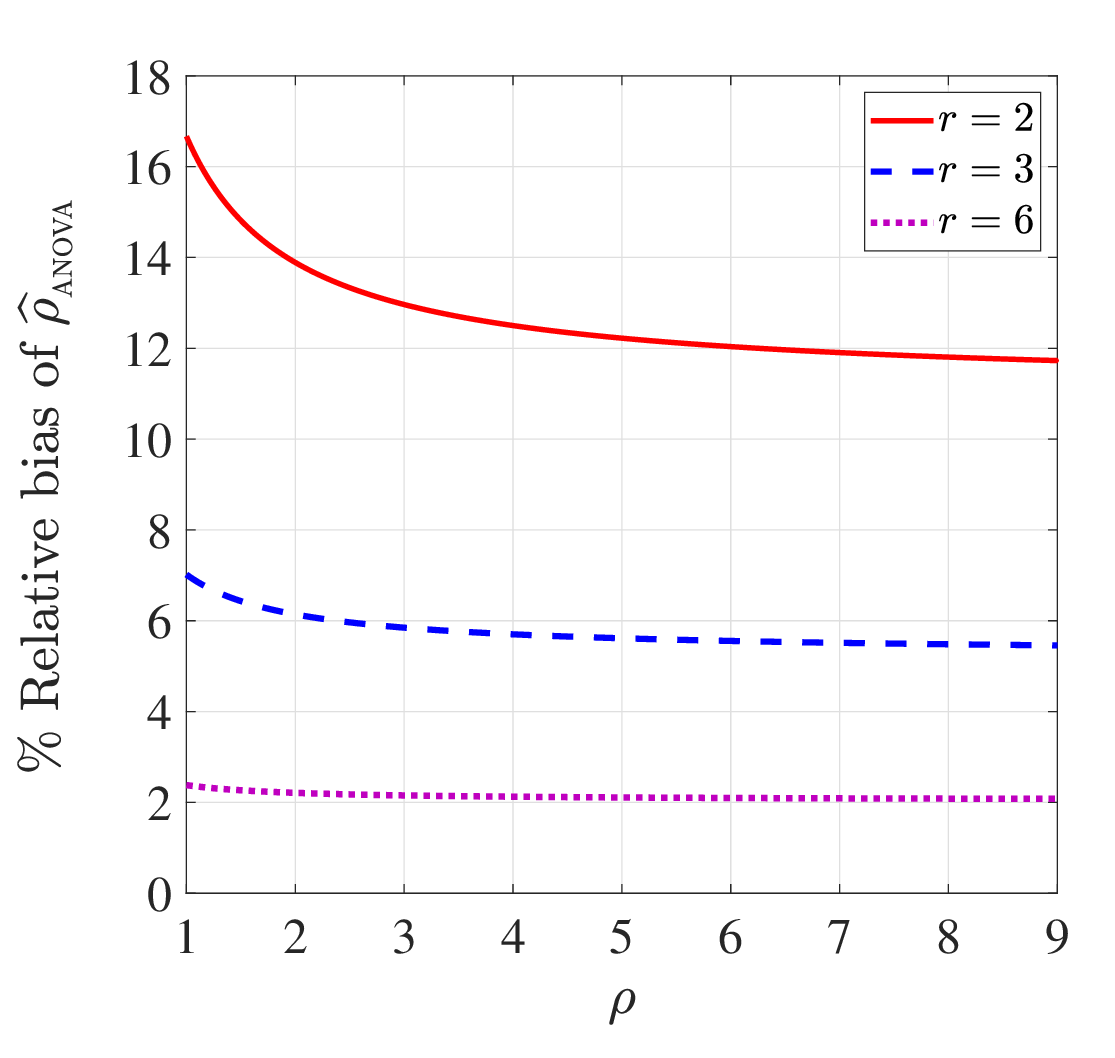}}}
\hspace{5pt}
\subfigure[$a=30$]{%
\resizebox*{50mm}{!}{\includegraphics{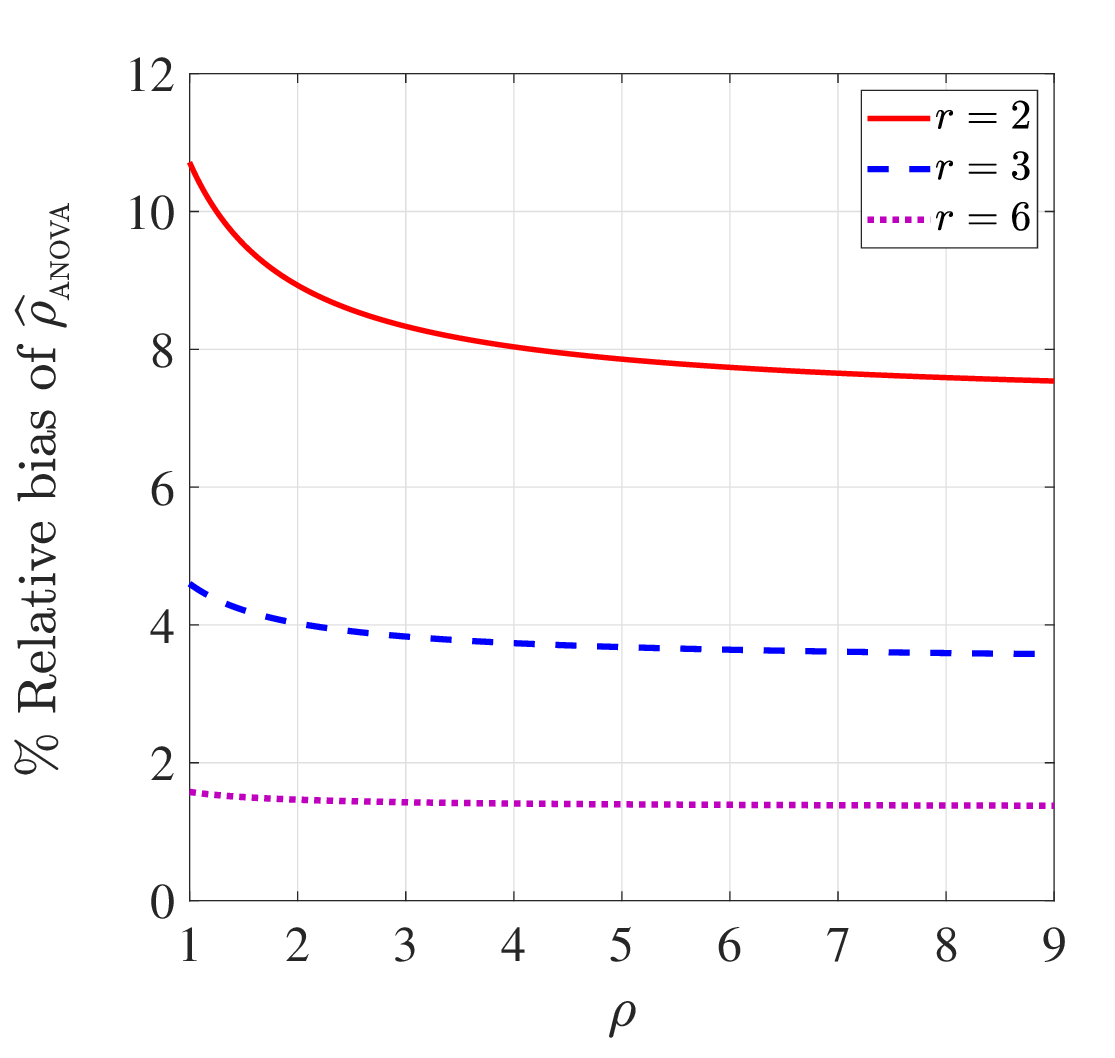}}}
\caption{The percentage relative bias of $\widehat{\rho}_{\ds{\tiny\mbox{ANOVA}}}$ with respect to $\rho$ in a one-way random effect model.}
\label{Fig:ANOVA-rho-bias} 
\end{figure}

The relative bias, expressed in percentage, provides a measure of an estimator's accuracy. For the estimator of $\rho$, it is calculated as
\begin{align}    
\left(\frac{\mathrm{E}\left[\widehat{\rho}\,\right]}{\rho}-1\right)\times 100 \,.
\end{align}
Figure~\ref{Fig:ANOVA-rho-bias} illustrates a comparison of the relative bias of  $\widehat{\rho}_{\ds{\tiny\mbox{ANOVA}}}$ with respect to $\rho$ for various study plans, characterized by different combinations of sample sizes and replication measurements.

It is worth noting that any reasonable measurement system typically exhibits \mbox{$\rho > 1$}~\citep{browne2010optimal}. Consequently, our study focuses on the range of $\rho > 1$, including four distinct sample sizes, $a=5, 10, 20, 30$. For each sample size,  three different numbers of replication measurements are considered, $r=2, 3, 6$. As shown in Figure~\ref{Fig:ANOVA-rho-bias}, increasing either the sample size or replications reduces the bias in the estimation of $\widehat{\rho}_{\ds{\tiny\mbox{ANOVA}}}$.

\subsubsection*{B. Maximum likelihood and non-negative ANOVA }

Identifying the bias of maximum likelihood and non-negative ANOVA estimators can be challenging since their estimations of $\sigma^2_u$ and $\sigma^2_\epsilon$ depend on the relationship between $MS_u$ and $MS_\epsilon$. Specifically, the expected values of $\sigma^2_u$, $\sigma^2_\epsilon$, and $\rho$ are influenced by the probability that the estimates lie on the boundary of the parameter space. This probability using the maximum likelihood estimation method is determined by, 
\begin{align}\label{eq:p_o}
    p_{_\circ} &=\Pr\left(\mfrac{MS_u}{MS_\epsilon} < \beta\right)  = \Pr\left(F < \beta{(1+r\rho)^{-1}}\right),
\end{align}
where $F$ is the random variable, as characterized in~\eqref{eq:F}. 
The expected values of the maximum likelihood estimators $\widehat{\sigma^2_{u}}$ and $\widehat{\sigma^2_\epsilon\,}$ are determined by
\begin{align}\label{eq:ch2-E-MLE-u}
    &\mathrm{E}\left[\widehat{\sigma^2_{u}}\right] = \mfrac{1}{r}\left(1-p_{_\circ}\right)\mathrm{E}\left[\beta^{-1}MS_u - MS_\epsilon \,\middle| \,\mfrac{MS_u}{MS_\epsilon} \geq \beta\right],\\
    &\mathrm{E}\left[\widehat{\sigma^2_\epsilon\,}\right] = \left(1-p_{_\circ}\right) \mathrm{E}\left[MS_\epsilon \middle| \,\mfrac{MS_u}{MS_\epsilon} \geq \beta\right]+  \frac{p_{_\circ}}{ar} \mathrm{E}\left[SS_t\middle| \,\mfrac{MS_u}{MS_\epsilon} < \beta\right],   \label{eq:ch2-E-MLE-e}
\end{align}
and the expected value of $\widehat{\rho}_{_{\tiny\mbox{MLE}}}$ is, 
\begin{align}\label{eq:E-MLE-rho}
    \mathrm{E}\left[\widehat{\rho}_{\ds{\tiny\mbox{MLE}}}\right] &= \mfrac{1}{r}\left(1-p_{_\circ}\right)\mathrm{E}\left[\beta^{-1}\mfrac{MS_u}{MS_\epsilon} - 1  \,\middle| \,\mfrac{MS_u}{MS_\epsilon} \geq \beta \right].
\end{align}
We can determine the expectation of $\widehat{\rho}_{_{\tiny\mbox{MLE}}}$ through,
\begin{align}\label{eq:ch2-E-MLE-rho}
    \mathrm{E}\left[\widehat{\rho}_{_{\tiny\mbox{MLE}}}\right] 
    &=\mfrac{1}{r}\left(1-p_{_\circ}\right)\mathrm{E}\left[\beta^{-1}(1+r\rho)F-1\mid F\geq \beta(1+r\rho)^{-1}\right].
\end{align}
The distribution of  $\widehat{\rho}_{_{\tiny\mbox{MLE}}}$ can be verified as a truncated $F$-distribution with $a-1$ and $a(r-1)$ degrees of freedom. A more explicit expression for the moments of this distribution is provided by~\cite{nadarajah2008moments}, which involves the use of the Gauss hypergeometric function.

The expected values of the maximum likelihood estimators $\widehat{\sigma^2_{u}}$, $\widehat{\sigma^2_\epsilon\,}$,  and $\widehat{\rho}_{_{\tiny\mbox{MLE}}}$  cannot be derived analytically due to the complexity of the underlying equations. As a result, determining the bias of these estimates analytically is challenging, necessitating numerical methods for evaluation.\smallskip

The properties of the estimators from the non-negative ANOVA and maximum likelihood methods are quite similar in model~\eqref{eq:one-way-model}. Specifically, by substituting $\beta$ with $1$ in equations~\eqref{eq:p_o} and~\eqref{eq:ch2-E-MLE-rho}, we can first determine the probability that the non-negative ANOVA estimations lie on the boundary and then compute $ \mathrm{E}\left[\widehat{\rho}_{\ds{\tiny\mbox{NANOVA}}}\right]$.\smallskip

\begin{figure}[!h]
\centering
\subfigure[$a=5$]{%
\resizebox*{50mm}{!}
{\includegraphics{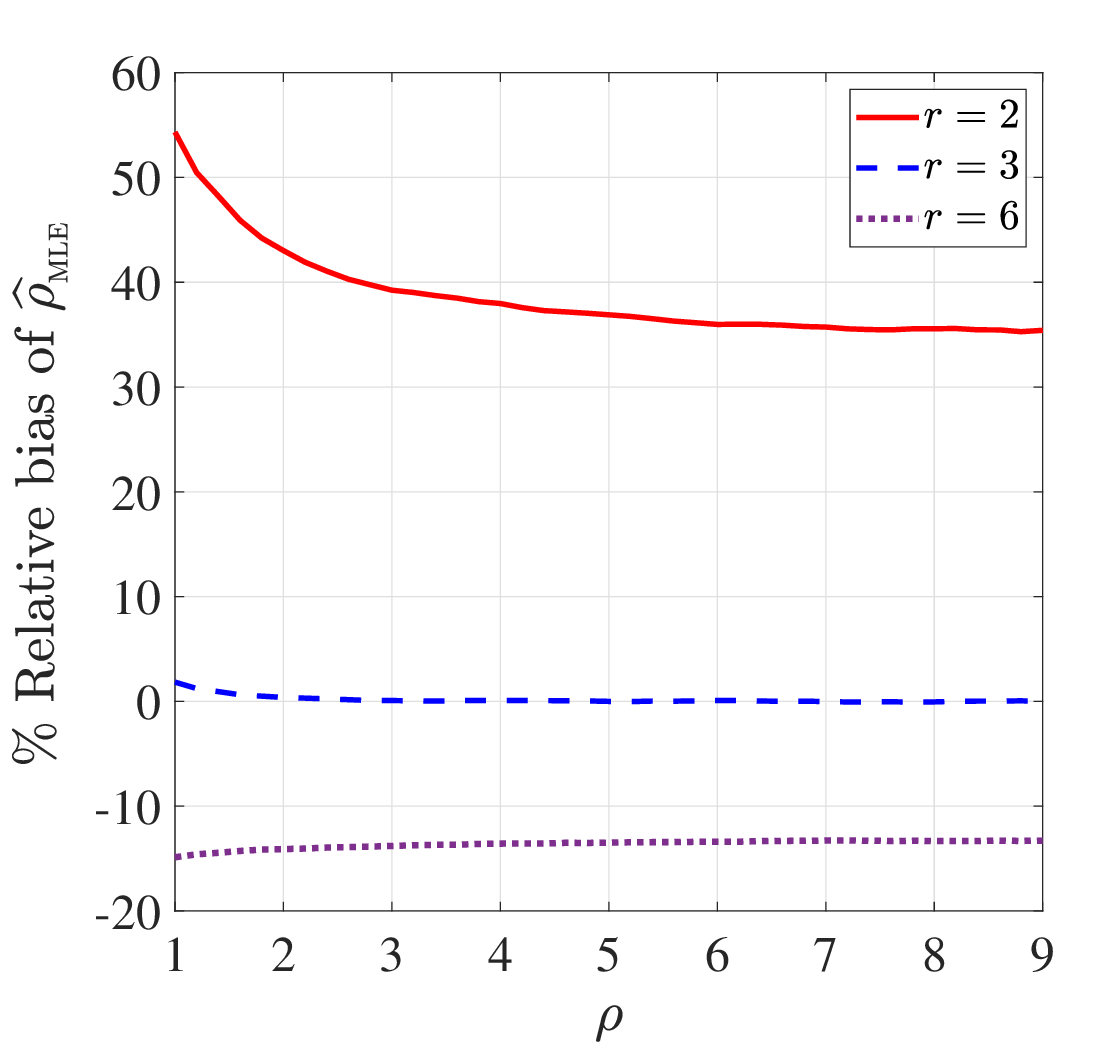}}}
\hspace{5pt}
\subfigure[$a=10$]{%
\resizebox*{50mm}{!}{\includegraphics{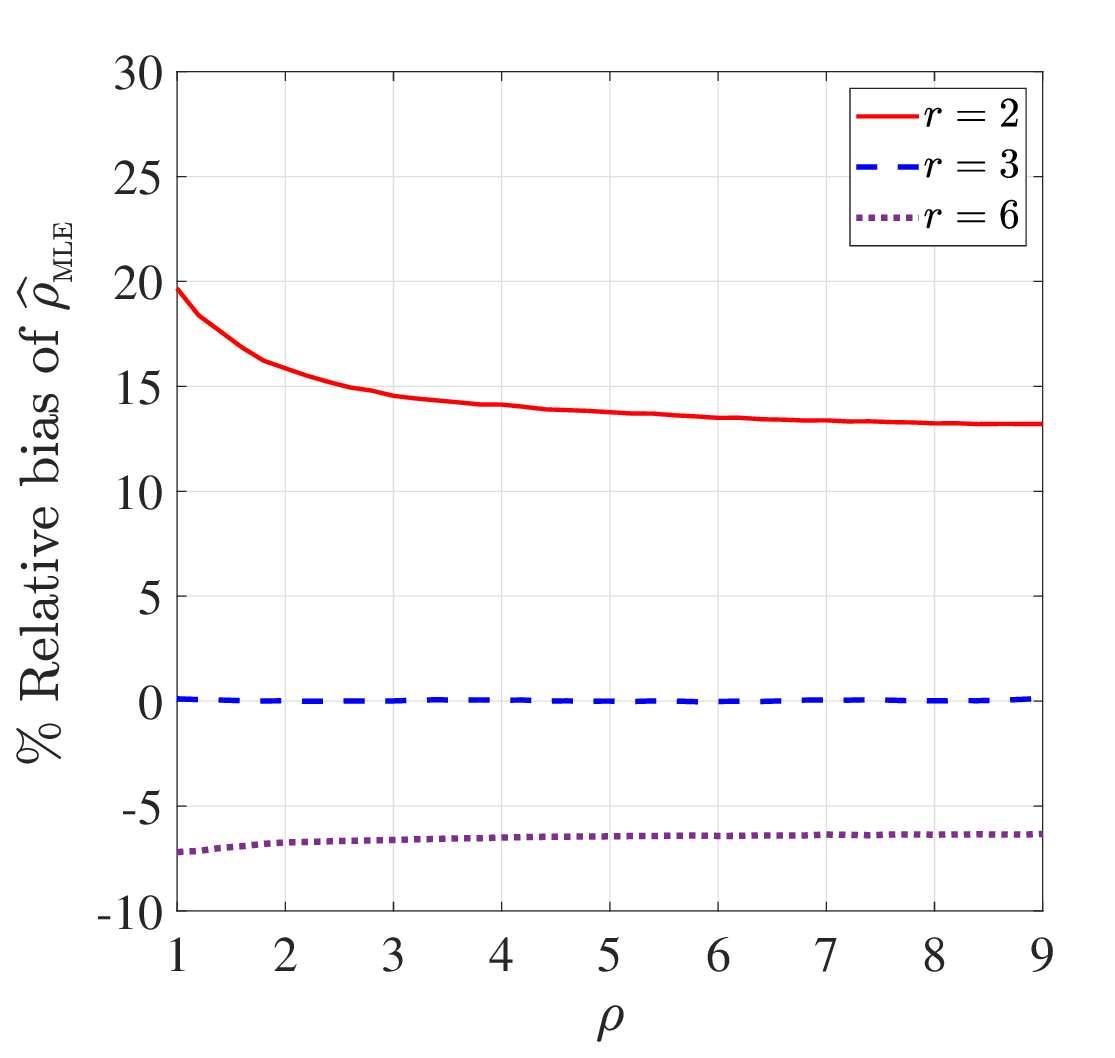}}}
\\
\subfigure[$a=20$]{%
\resizebox*{50mm}{!}{\includegraphics{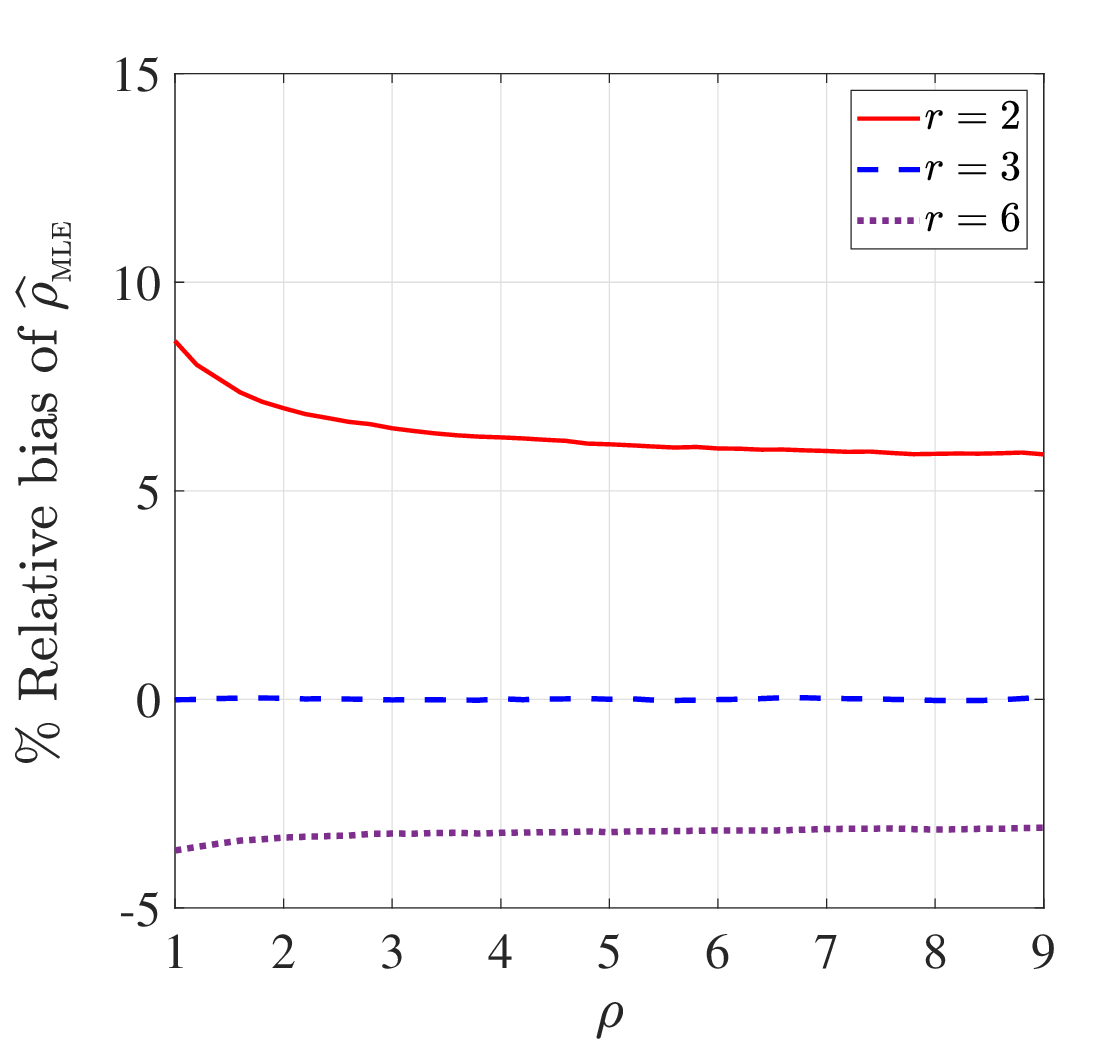}}}
\hspace{5pt}
\subfigure[$a=30$]{%
\resizebox*{50mm}{!}{\includegraphics{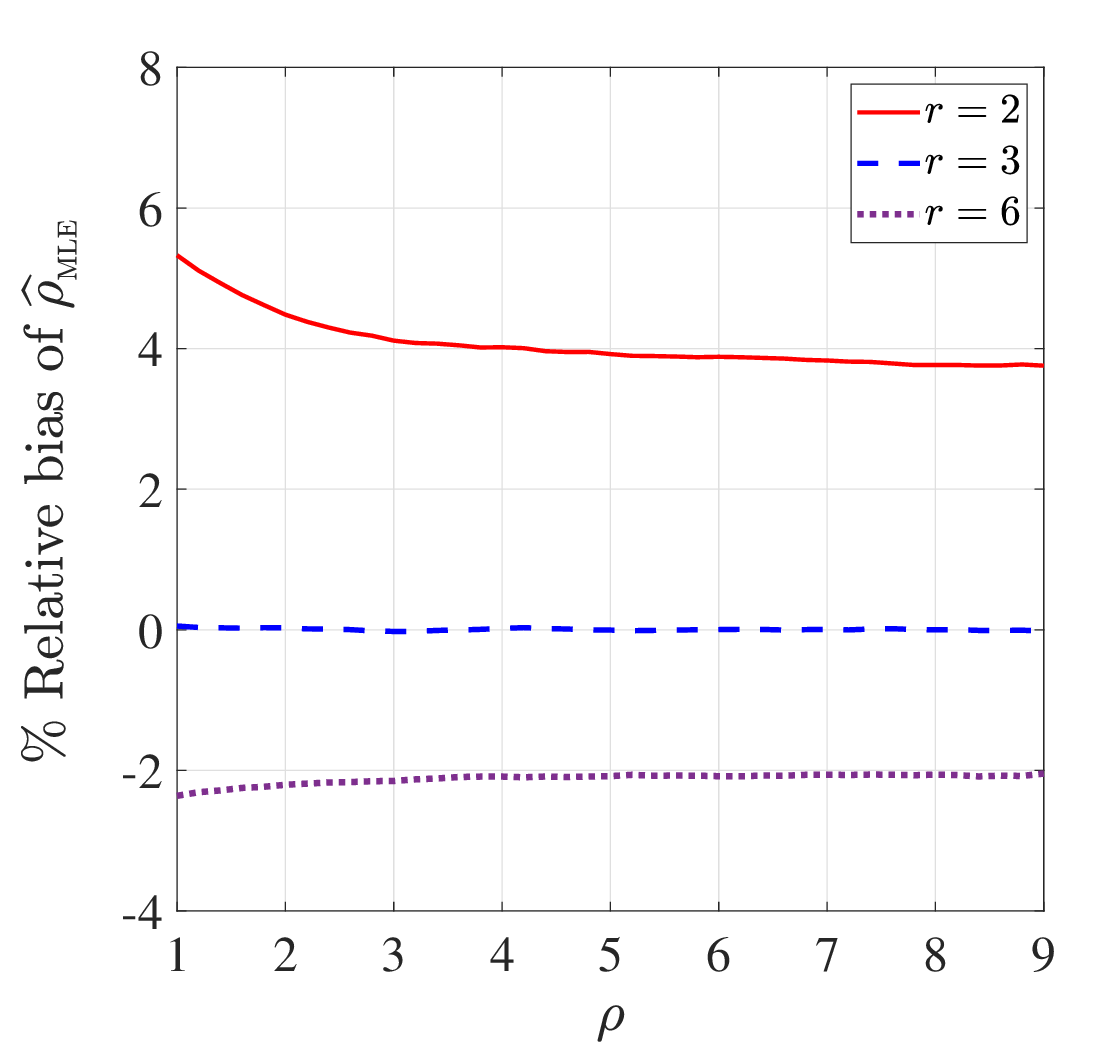}}}
\caption{The percentage relative bias of $\widehat{\rho}_{_{\tiny\mbox{MLE}}}$ with respect to $\rho$ in a one-way random effect model, where the random effects of $U$ and $\epsilon$ are normally distributed.}
\label{Fig:MLE-rho-bias} 
\end{figure}

\begin{figure}[!h]
\centering
\subfigure[$a=5$]{%
\resizebox*{50mm}{!}
{\includegraphics{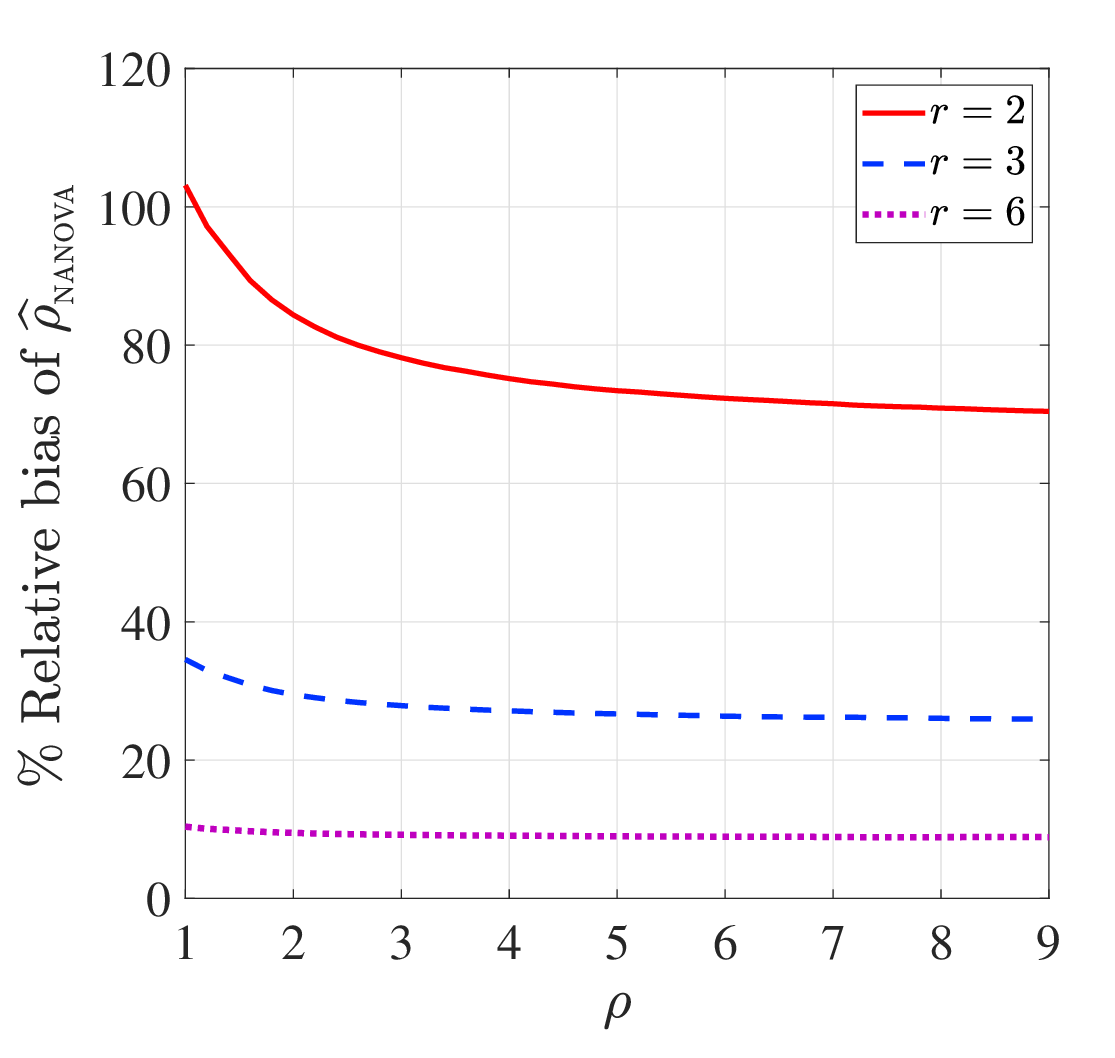}}}
\hspace{5pt}
\subfigure[$a=10$]{%
\resizebox*{50mm}{!}{\includegraphics{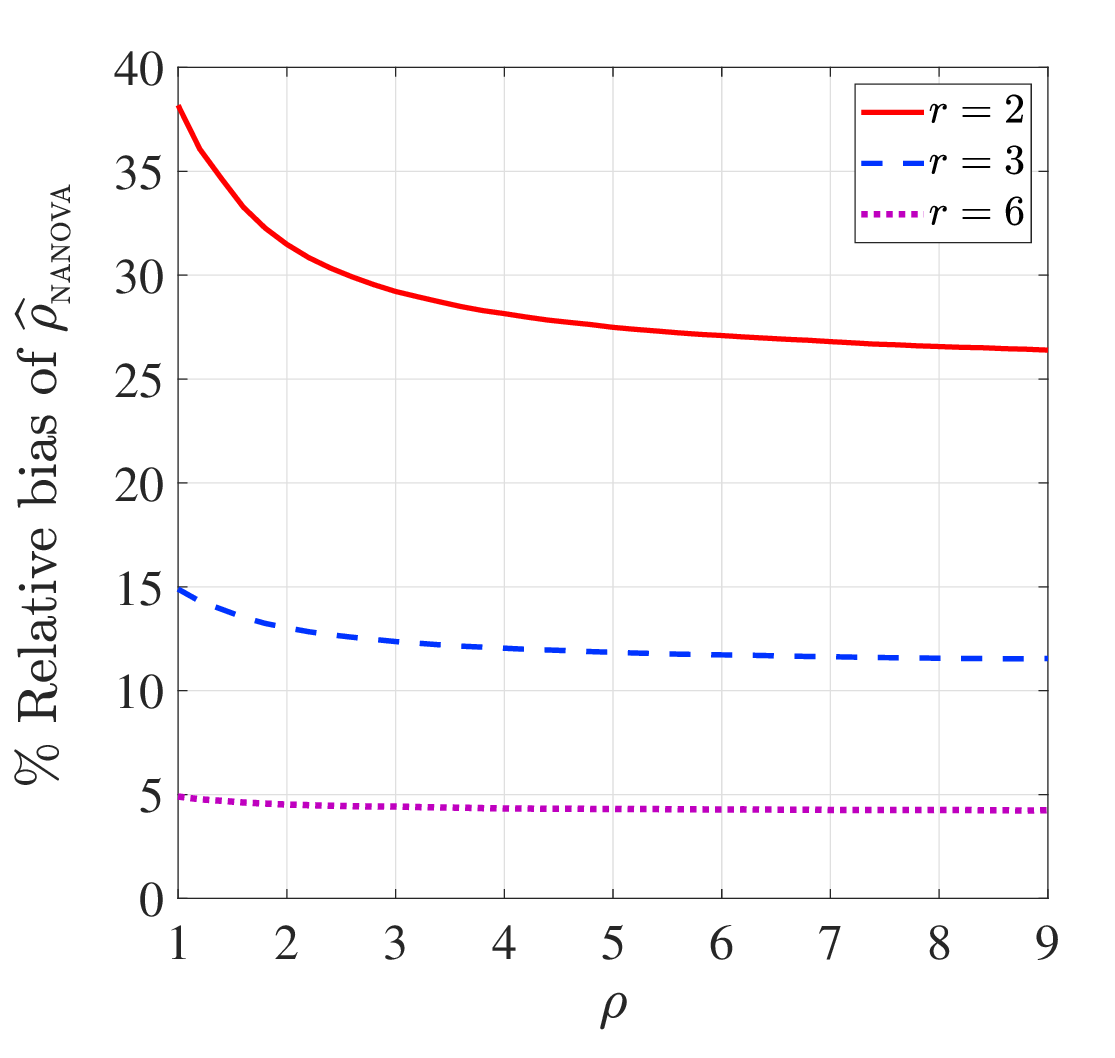}}}
\\
\subfigure[$a=20$]{%
\resizebox*{50mm}{!}{\includegraphics{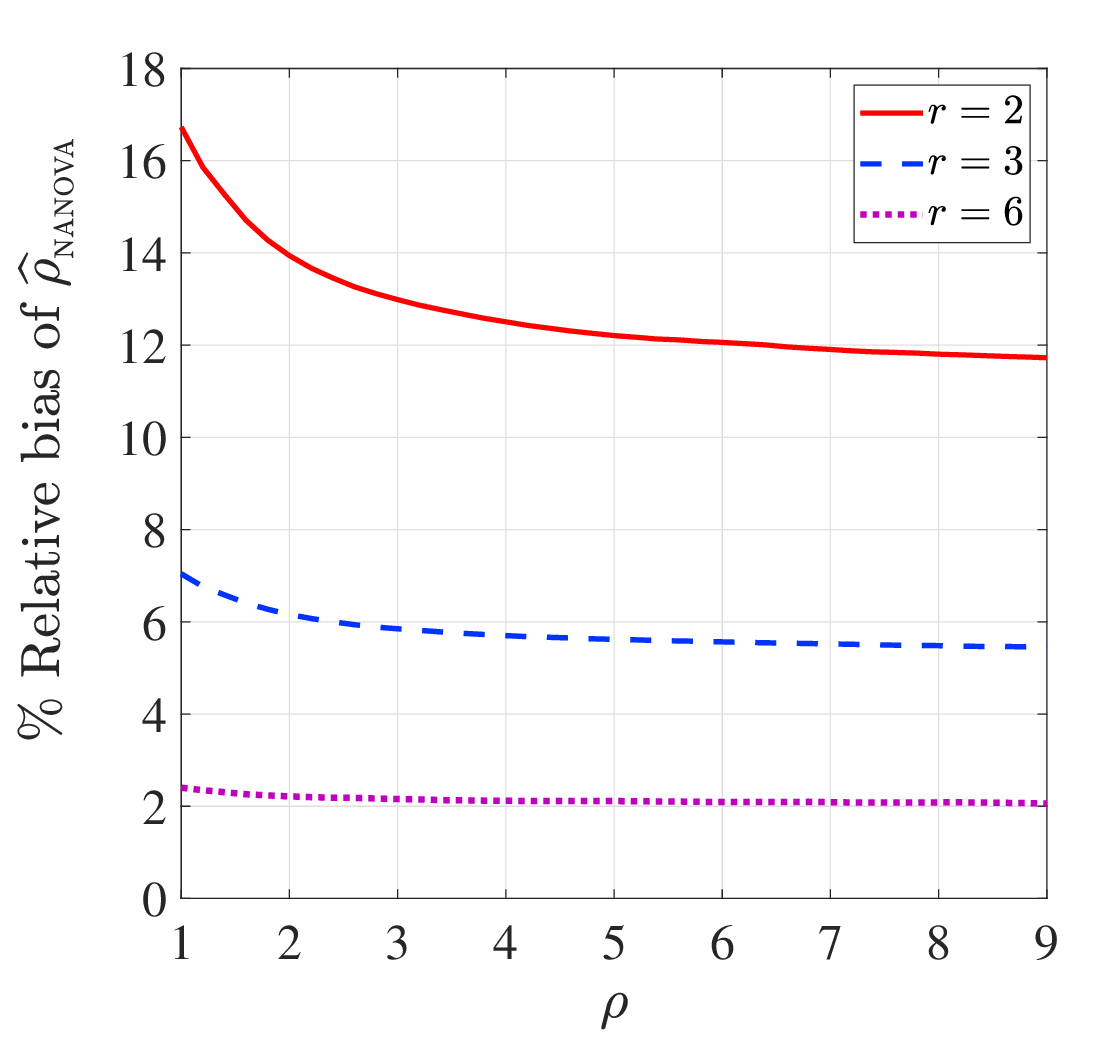}}}
\hspace{5pt}
\subfigure[$a=30$]{%
\resizebox*{50mm}{!}{\includegraphics{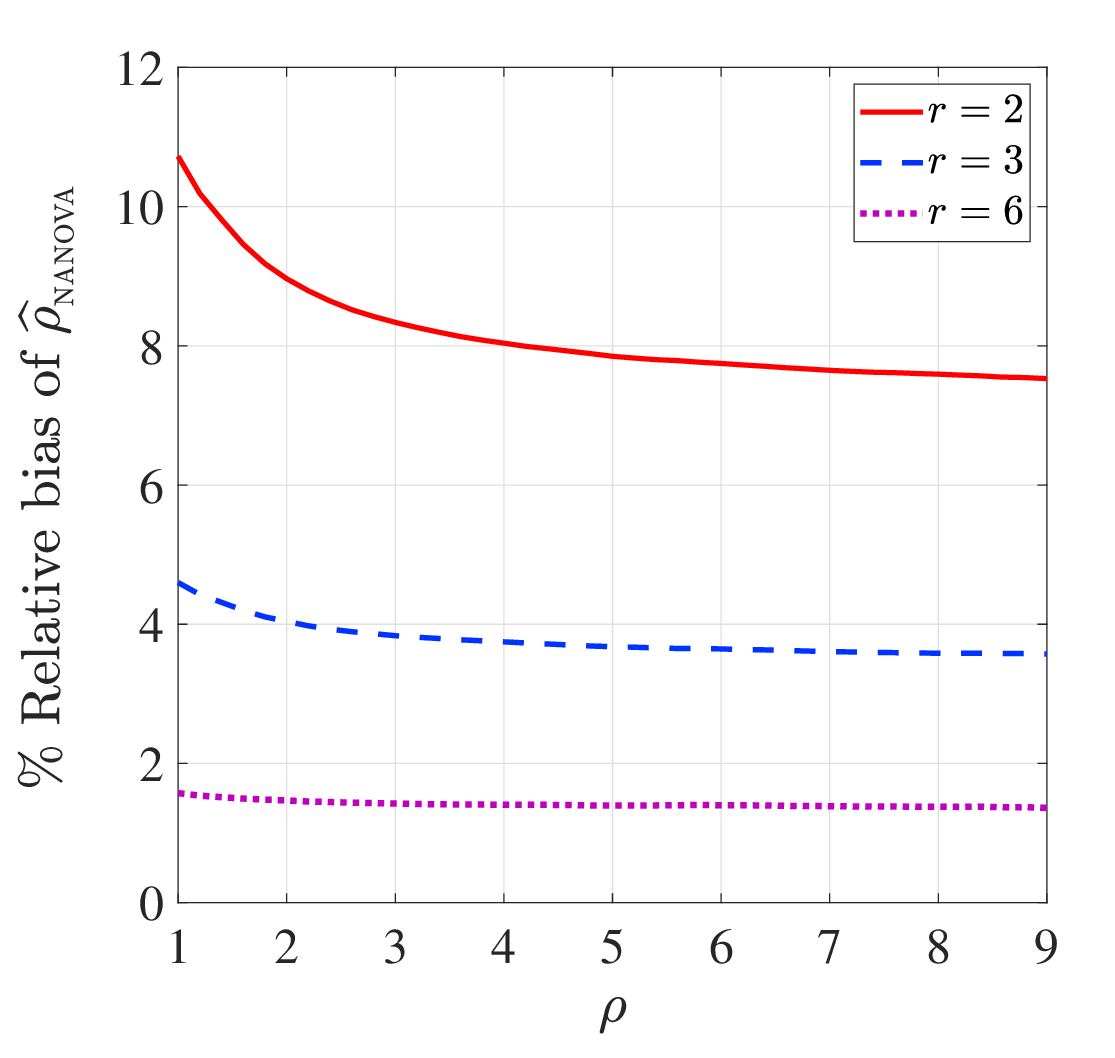}}}
\caption{The percentage relative bias of $\widehat{\rho}_{_{\tiny\mbox{NANOVA}}}$ with respect to $\rho$ in a one-way random effect model, where the random effects of $U$ and $\epsilon$ are normally distributed.}
\label{Fig:NANOVA-rho-bias} 
\end{figure}

Figures~\ref{Fig:MLE-rho-bias} and~\ref{Fig:NANOVA-rho-bias} display the percentage relative bias of $\widehat{\rho}_{_{\tiny\mbox{MLE}}}$ and $\widehat{\rho}_{_{\tiny\mbox{ANOVA}}}$ relative to $\rho$ across various combinations of $a$ and $r$, as shown in Figure~\ref{Fig:ANOVA-rho-bias}. From Figure~\ref{Fig:MLE-rho-bias}, the estimate of $\widehat{\rho}_{_{\tiny\mbox{MLE}}}$ exhibits a positive bias when $r=2$, while it shows a negative bias when $r=6$. Interestingly, when $r=3$, the bias remains approximately zero across all sample sizes, which is a desirable characteristic. In plot (d) of Figure~\ref{Fig:MLE-rho-bias} with $a=30$, we observe that the larger sample size results in a narrower range of relative bias, indicating reduced variability across different numbers of measurement replications. Comparing Figures~\ref{Fig:MLE-rho-bias} and~\ref{Fig:NANOVA-rho-bias}, we find that $\widehat{\rho}_{_{\tiny\mbox{MLE}}}$ generally exhibits smaller bias for plans with fewer measurement replications (e.g., $r=2$ and $r=3$). However, $\widehat{\rho}_{_{\tiny\mbox{NANOVA}}}$ has a smaller relative bias than $\widehat{\rho}_{_{\tiny\mbox{MLE}}}$, in absolute terms, when $r=6$.

\begin{figure}[!h]
\centering
\subfigure[ANOVA]{%
\resizebox*{46mm}{!}{\includegraphics{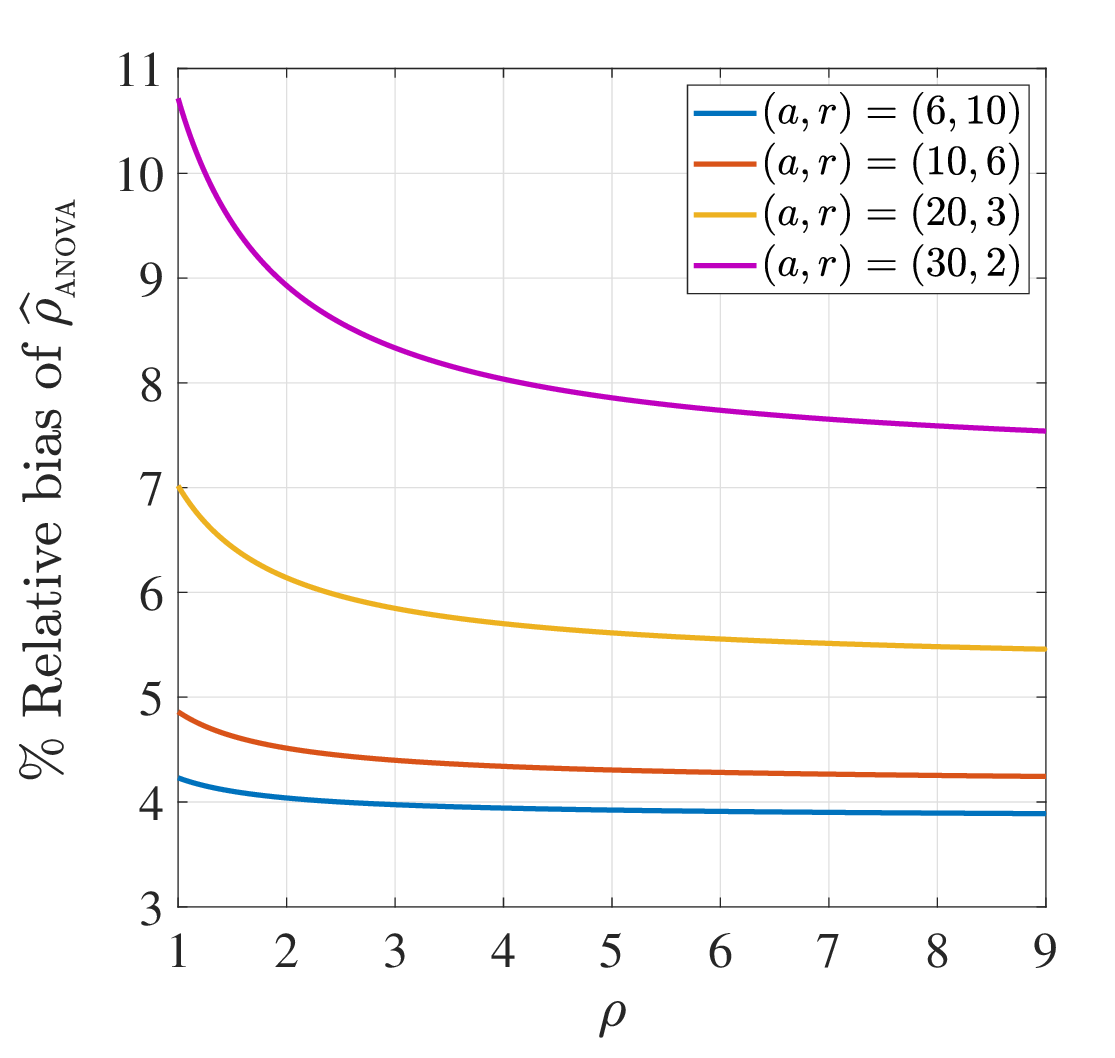}}}
\hspace{1pt}
\subfigure[Maximum likelihood]{%
\resizebox*{46mm}{!}
{\includegraphics{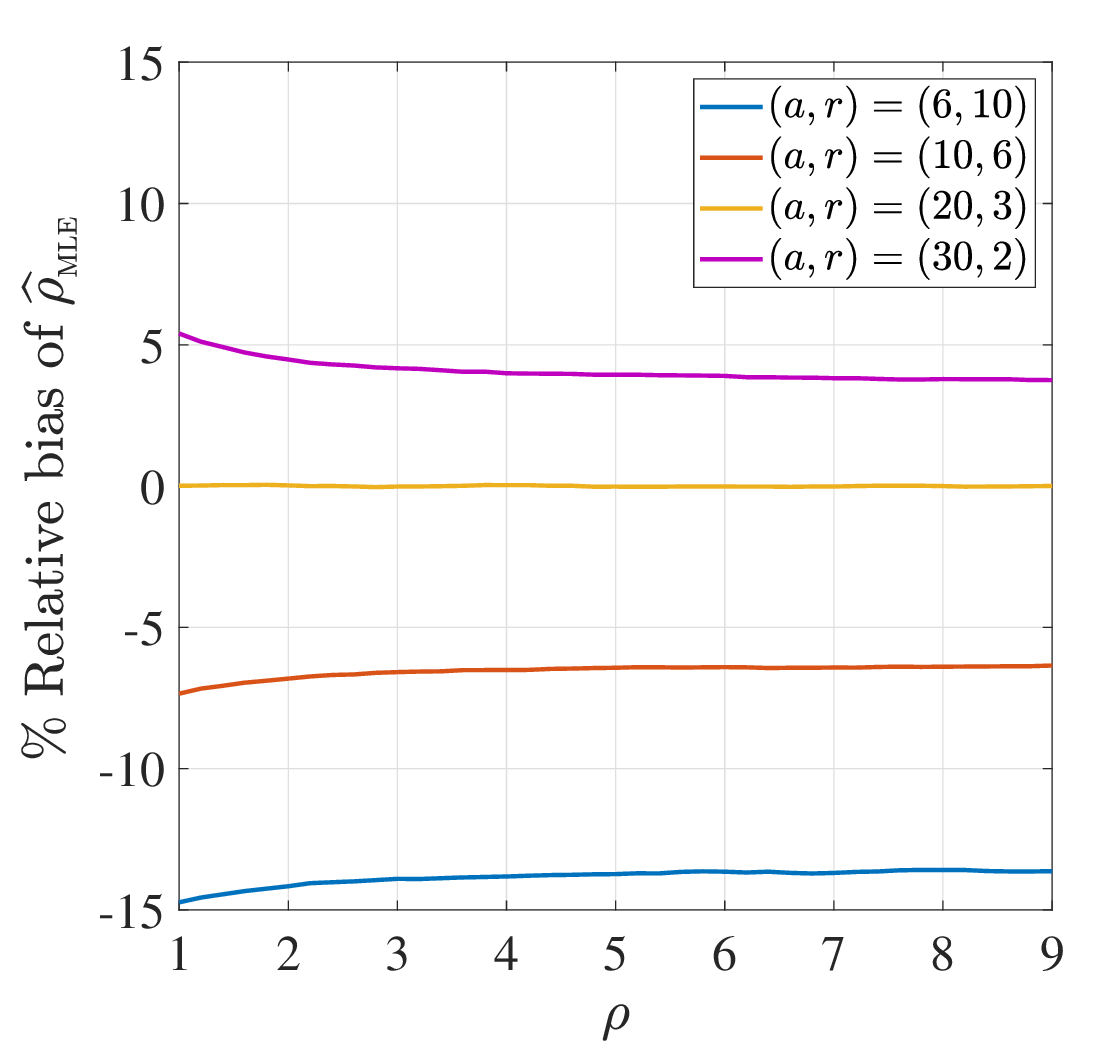}}}
\hspace{1pt}
\subfigure[Non-negative ANOVA]{%
\resizebox*{46mm}{!}{\includegraphics{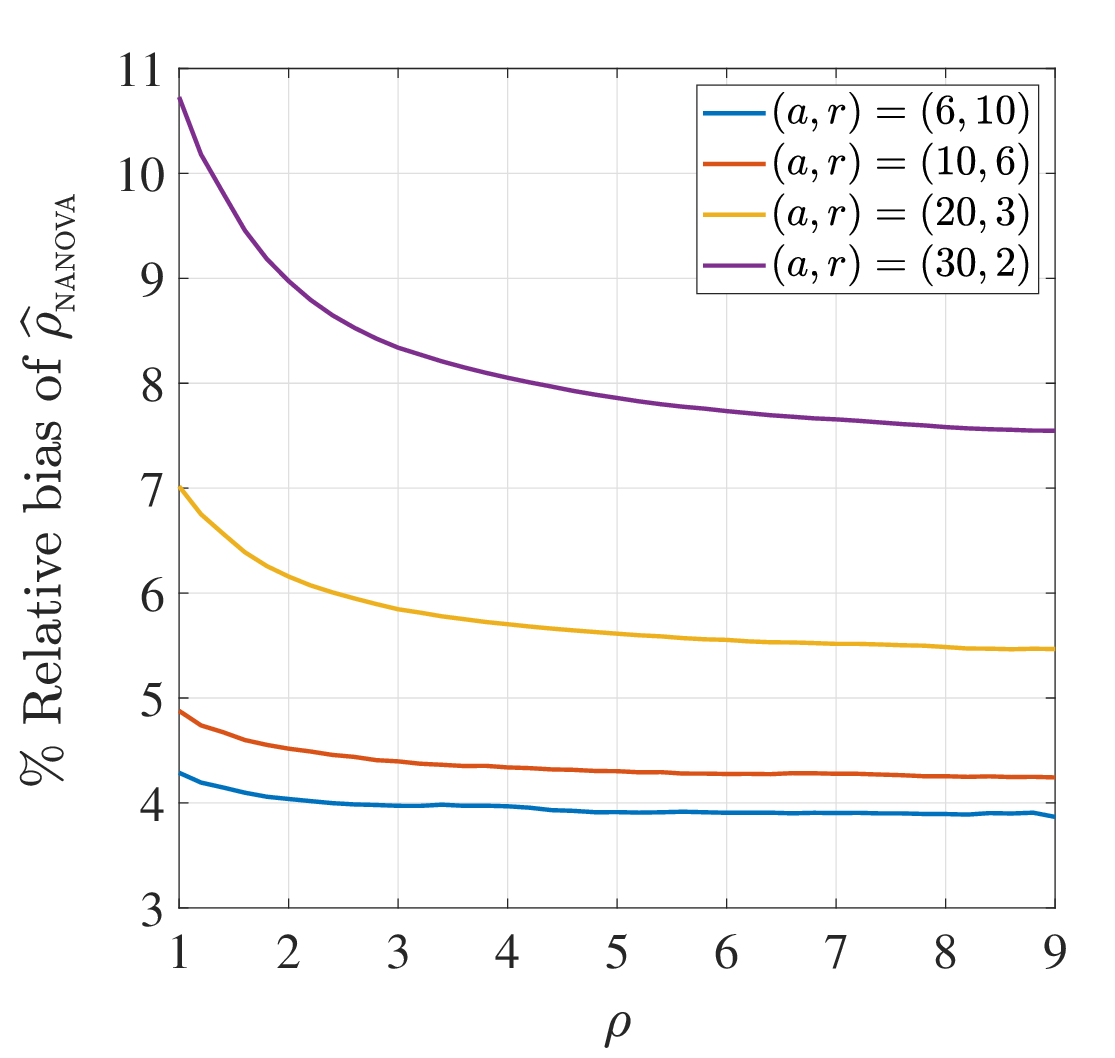}}}
\caption{The percentage relative bias of (a) $\widehat{\rho}_{\ds{\tiny\mbox{ANOVA}}}$, (b) $\widehat{\rho}_{_{\tiny\mbox{MLE}}}$ and (c) $\widehat{\rho}_{\ds{\tiny\mbox{NANOVA}}}$ as a function of $\rho$ in a one-way random effect model with $N=60$ measurements.}
\label{Fig:Bias_rho_MLE_NANOVA-FixedN} 
\end{figure}

Figure~\ref{Fig:Bias_rho_MLE_NANOVA-FixedN} illustrates the percentage relative bias of $\widehat{\rho}_{\ds{\tiny\mbox{ANOVA}}}$ , $\widehat{\rho}_{_{\tiny\mbox{MLE}}}$ and $\widehat{\rho}_{\ds{\tiny\mbox{NANOVA}}}$ across 
four study plans, each with a constant total number of measurements $N=60$. We note that $\widehat{\rho}_{_{\tiny\mbox{MLE}}}$ demonstrates negligible bias, approaching zero, in the plan with $a=20$ and $r=3$. In contrast, $\widehat{\rho}_{\ds{\tiny\mbox{ANOVA}}}$ and $\widehat{\rho}_{\ds{\tiny\mbox{NANOVA}}}$ achieve their lowest bias in the plan with $a=6$ and $r=10$.

\subsection{Sampling variance}
\subsubsection*{A.~ANOVA estimations}
The variances of the mean of squares terms $MS_u$ and $MS_\epsilon\,$, which are proportional to chi-squared distributions with degrees of freedom $\mathrm{df}_u = a-1$ and $\mathrm{df}_\epsilon = a(r-1)$, are given as follows~\citep{searle1992variance}, 
\begin{align}
\Var\left[MS_u\right] = \frac{2\left(\mathrm{E}\left[MS_u\right]\right)^2}{a-1}
    \quad
\mbox{and }\quad \Var\left[MS_\epsilon\right]=\frac{2\left(\mathrm{E}\left[MS_\epsilon\right]\right)^2}{a(r-1)}\,\cdot
\end{align}
Assuming the independence of $MS_u$ and $MS_\epsilon$ under normality, the variances of $\widehat{\sigma^2_{u}}$ and $\widehat{\sigma^2_\epsilon,}$ using the ANOVA estimation method, for a sample size of $a$ and replication $r$ can be expressed as follows,
\begin{align}
    \Var\left[\widehat{\sigma^2_u}\right]&= \mfrac{2}{r^2}\left(\frac{(\sigma^2_\epsilon + r \sigma^2_u)^2}{a-1}+\frac{\sigma^4_\epsilon}{a(r-1)}\right), \\
    \Var\left[\widehat{\sigma^2_{\epsilon\,}}\right] &= \frac{2\,\sigma^4_\epsilon}{a(r-1)}.
\end{align}
The variance of $\widehat{\rho}_{\ds{\tiny\mbox{ANOVA}}}$, as defined in~\eqref{eq:ANOVA-rho}, can be computed by,
\begin{align} \label{eq:ch2-rho_ANOVA-var}      \Var\left[\widehat{\rho}_{\ds{\tiny\mbox{ANOVA}}}\right]  &=\mfrac{1}{r^2}\,\mathrm{Var}\left[\frac{MS_u}{MS_\epsilon}\right] \nonumber\\
    &=\frac{(1+r\rho)^2}{r^2}\,\mathrm{Var}\left[F\right],
\end{align}
where the random variable $F$ is defined in~\eqref{eq:F}. The variance of random variable $F$ is,
\begin{align}\label{eq:ch2-F-var}
    \mathrm{Var}\left[F\right] 
    &=\frac{2\,\mathrm{df}_\epsilon^2\left(\mathrm{df}_\epsilon+\mathrm{df}_u-2\right)}{\mathrm{df}_u\left(\mathrm{df}_\epsilon-2\right)^2\left(\mathrm{df}_\epsilon-4\right)},
\end{align}
where $\mathrm{df}_\epsilon>4$, and the variance does not exist otherwise. By substituting \eqref{eq:ch2-F-var} in to \eqref{eq:ch2-rho_ANOVA-var}, the variance of $\widehat{\rho}_{\ds{\tiny\mbox{ANOVA}}}$ is obtained as,
\begin{align}
    \mathrm{Var}\left[\widehat{\rho}_{\ds{\tiny\mbox{ANOVA}}}\right] 
    &=\frac{2(1+r\rho)^2\,\mathrm{df}_\epsilon^2\left(\mathrm{df}_\epsilon+\mathrm{df}_u-2\right)}{r^2\mathrm{df}_u\left(\mathrm{df}_\epsilon-2\right)^2\left(\mathrm{df}_\epsilon-4\right)}.
\end{align}

\subsubsection*{B.~Maximum likelihood and non-negative ANOVA}
First, we analyze the variance of $\widehat{\sigma^2_{u}}$ and $\widehat{\sigma^2_\epsilon\,}$ using their maximum likelihood estimations. For $\widehat{\sigma^2_{u}}$, the variance is given by,
\begin{align}\label{eq:Var_MLE_U}
    \mathrm{Var}\left[\widehat{\sigma^2_{u}}\right] &=\mathrm{E}\left[\big(\,\widehat{\sigma^2_u}\,\big)^2\right]-\left(\mathrm{E}\left[\widehat{\sigma^2_u}\right] \right)^2 \nonumber\\
    &= \mfrac{1}{r^2}(1-p_{_\circ})\mathrm{E}\left[\left(\beta^{-1}MS_u - MS_\epsilon\right)^2 \,\middle| \,\mfrac{MS_u}{MS_\epsilon} \geq \beta\right] - \left(\mathrm{E}\left[\widehat{\sigma^2_u}\right] \right)^2,
\end{align}
where $\mathrm{E}\left[\widehat{\sigma^2_u}\right]$ is provided in \eqref{eq:ch2-E-MLE-u}. \smallskip
For $\widehat{\sigma^2_\epsilon\,}$, the variance is expressed as, 
\begin{align}\label{eq:Var_MLE_E}
    \!\!\mathrm{Var}\left[\widehat{\sigma^2_\epsilon\,}\right] &=\mathrm{E}\left[\big(\,\widehat{\sigma^2_\epsilon\,}\,\big)^2\right]-\left(\mathrm{E}\left[\widehat{\sigma^2_\epsilon\,}\right] \right)^2 \nonumber\\
    &= (1-p_{_\circ})\mathrm{E}\left[MS_\epsilon^2 \,\middle| \,\mfrac{MS_u}{MS_\epsilon} \geq \beta\right] +  \mfrac{p_{_\circ}}{a^2r^2}\, \mathrm{E}\left[SS_t^2\middle| \,\mfrac{MS_u}{MS_\epsilon} < \beta\right]- \left(\mathrm{E}\left[\widehat{\sigma^2_\epsilon}\right] \right)^2.
\end{align}
where $\mathrm{E}\left[\widehat{\sigma^2_\epsilon}\right]$ is detailed in~\eqref{eq:ch2-E-MLE-e}. Similarly, the variance of $\widehat{\rho}_{_{\tiny\mbox{MLE}}}$ can be determined as 
\begin{align}
    \mathrm{Var}\left[\widehat{\rho}_{_{\tiny\mbox{MLE}}}\right] =\mathrm{E}\left[\widehat{\rho\,}_{\ds{\tiny\mbox{MLE}}}^2\right]-\left(\mathrm{E}\left[\,\widehat{\rho}_{_{\tiny\mbox{MLE}}}\right] \right)^2, \nonumber
\end{align} 
where, 
\begin{align}
    \mathrm{E}\left[\widehat{\rho\,}_{\ds{\tiny\mbox{MLE}}}^2\right]      &=\mfrac{1}{r^2}(1-p_{_\circ})\mathrm{E}\left[\left(\beta^{-1}(1+r\rho)F-1\right)^2\mid F\geq \beta(1+r\rho)^{-1}\right],
\end{align}
and $\mathrm{E}\left[\widehat{\rho}_{_{\tiny\mbox{MLE}}}\right] $ is defined in~\eqref{eq:ch2-E-MLE-rho}.

The finite-sample variance properties of  $\widehat{\sigma^2_{u}}$,  $\widehat{\sigma^2_\epsilon\,}$ and $\widehat{\rho}_{\ds{\tiny\mbox{NANOVA}}}$, using non-negative ANOVA estimators, can be derived in a similar manner by setting $\beta$ to $1$. 

To compare the finite-sample variances of $\rho$ estimations, a relative standard error (SE) is calculated and expressed as a percentage using the formula
\begin{align}
    \frac{\sqrt{\mathrm{Var\left[\,\widehat{\rho}\,\right]}}}{\rho}\times 100\,.
\end{align}

\begin{figure}[!h]
\centering
\subfigure[Maximum likelihood]{%
\resizebox*{50mm}{!}
{\includegraphics{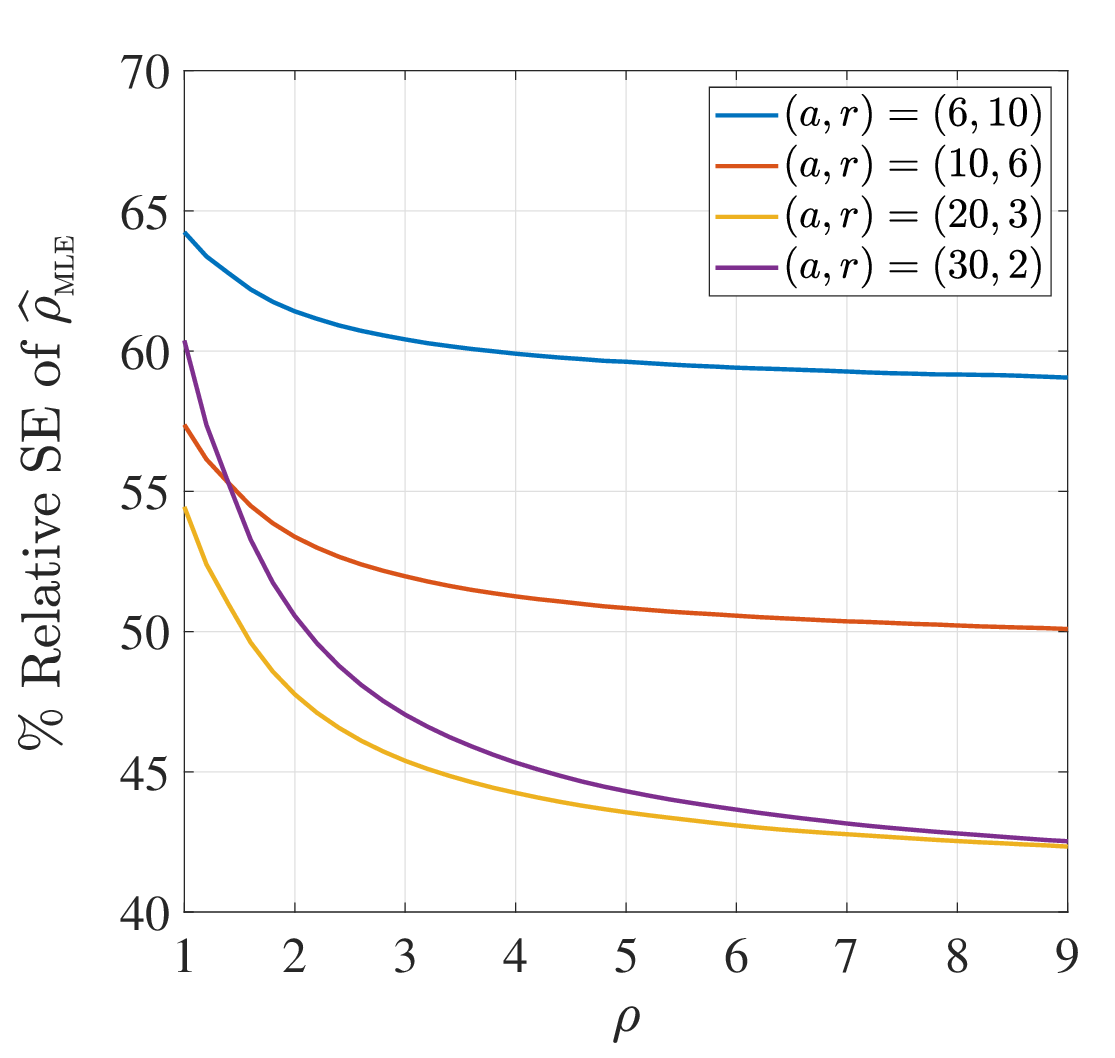}}}
\hspace{10pt}
\subfigure[Nan-negative ANOVA]{%
\resizebox*{50mm}{!}{\includegraphics{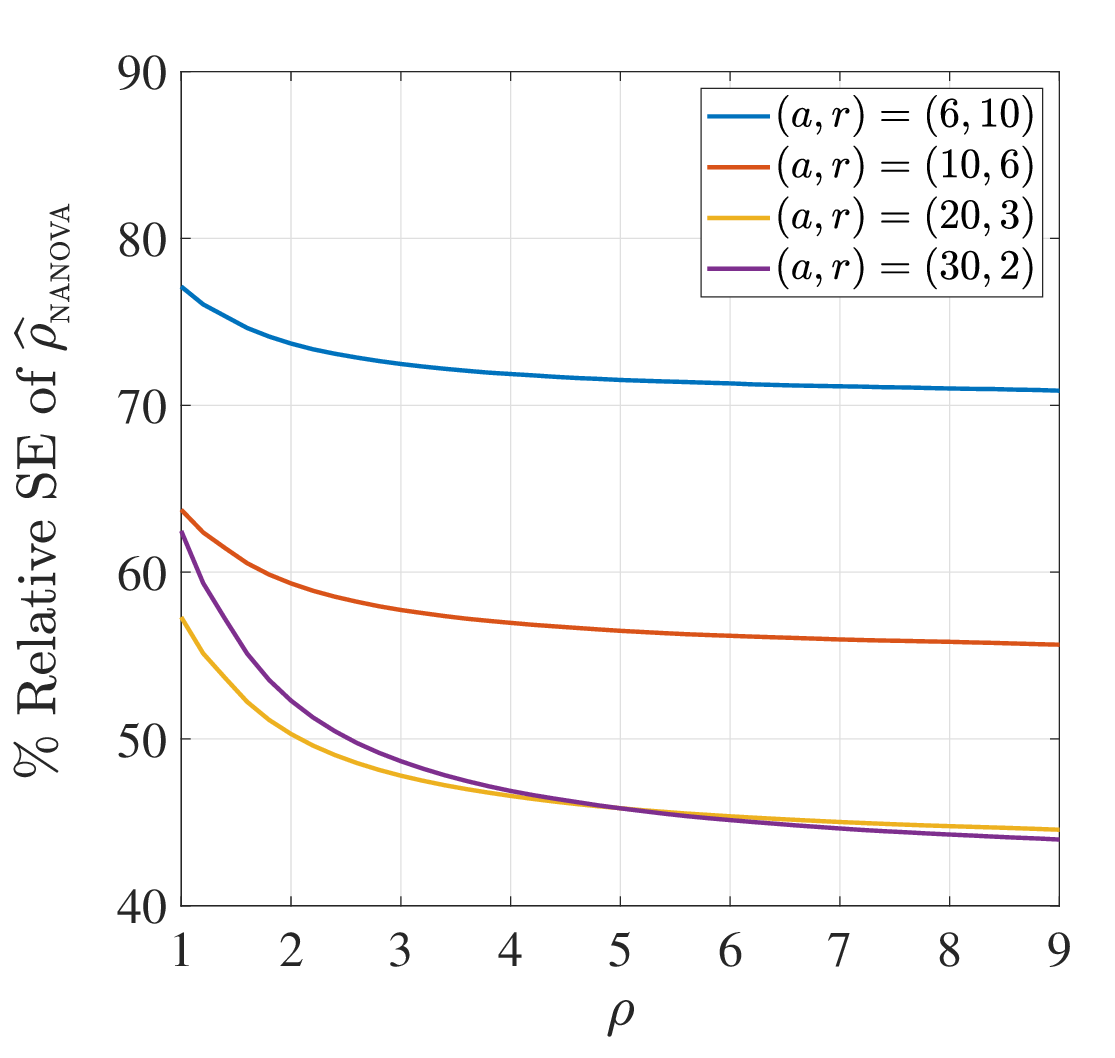}}}
\caption{The percentage relative SE of (a) $\widehat{\rho}_{_{\tiny\mbox{MLE}}}$ and (b) $\widehat{\rho}_{_{\tiny\mbox{NANOVA}}}$ as a function of $\rho$ in a one-way random effect model with $N=60$ measurements.}
\label{Fig:SE_rho_MLE_NANOVA-FixedN} 
\end{figure}

Figure~\ref{Fig:SE_rho_MLE_NANOVA-FixedN} illustrates the relative standard error (SE) as a function of $\rho$ for the estimates $\widehat{\rho}_{\ds{\tiny\mbox{NANOVA}}}$ and $\widehat{\rho}_{\ds{\tiny\mbox{MLE}}}$ under various study plans, each constrained to a total of $60$ measurements. The comparison reveals that $\widehat{\rho}_{\ds{\tiny\mbox{MLE}}}$ consistently exhibits lower variance than $\widehat{\rho}_{\ds{\tiny\mbox{NANOVA}}}$ across the examined plans. Notably, the smallest relative SE for $\widehat{\rho}_{\ds{\tiny\mbox{MLE}}}$ is observved when the study plan includes a sample size of $20$ with $3$ measurement replications.

\subsection{Test of hypothesis}\label{sec:hypothesis-test}
Point estimates provide valuable insight into the magnitude of assessment parameters in a measurement system study. However, they offer limited information regarding the precision of these estimates. For instance, if the estimated variability of the measurement system exceeds expectations, hypothesis testing can determine whether it is significantly larger than a specified threshold. Similarly, a small point estimate for the signal-to-noise ratio (or the parameter $\rho$) may suggest an inadequate measurement system. Statistical hypothesis tests allow for probabilistic evaluations, enabling the assessment of the significance of these parameters \citep{burdick1994using,montgomery1993gaugeI}.

To evaluate the statistical significance of estimated values and facilitate informed decision-making regarding the measurement system, we will examine three specific hypothesis tests within the framework of model~\eqref{eq:one-way-model}, as outlined below.
\begin{enumerate}[leftmargin=2.0\parindent,label=\roman*)]
    \item $H_{_0}\,:\, \sigma^2_u = 0$ vs $H_{_0}\,:\, \sigma^2_u > 0\,$,
    \item $H_{_0}\,:\,\sigma_\epsilon \leq \sigma_{_0}$ vs $H_{a}\,:\, \sigma_\epsilon > \sigma_{_0}\,$,
    \item $H_{_0}\,:\, \rho \leq \rho_{_0}$ vs  $H_{a}\,:\,\rho >\rho_{_0}\,$.
\end{enumerate}
An essential question in measurement system analysis is whether variability in measurement data can be attributed to differences among units \citep{montgomery1993gaugeI}. To investigate this, a hypothesis test can be conducted for $\sigma^2_u$ with the null hypothesis $H_{_0}\,:\, \sigma^2_u = 0\,$.

We use the statistic ${MS_u}/{MS_\epsilon}$ to  assess whether the variance component $\sigma^2_u$ is negligible compared to $\sigma^2_\epsilon$. Under hypothesis $H_{_0}\,:\, \sigma^2_u = 0$, the statistic ${MS_u}/{MS_\epsilon}$ follows an $F$-distribution with $\mathrm{df}_u$ and $\mathrm{df}_\epsilon$ degrees of freedom. A large observed value of this statistic provides evidence that the variance component $\sigma^2_u$ significantly contributes to the variability in the data. The $p$-value for testing $H_{_0}\,:\, \sigma^2_u = 0$ is calculated as follows, 
\begin{align}\label{eq:uni-p-value}
    p\mbox{-value} = 1- \Pr\left(F < \frac{r\sum_{i=1}^a (y_{i\cdot} - \overline{y}_{\cdot\cdot})^2/(a-1)}{\sum_{i=1}^a\sum_{j=1}^r (y_{ij} - \overline{y}_{i\cdot})^2/(N-a)}\right), 
\end{align} 

A frequent question in measurement system assessment studies is whether the variability of the measurement system remains below an acceptable threshold~\citep{burdick1997confidence}. To address this, a hypothesis test can be conducted with the null hypothesis $H_{_0}\,:\,\sigma_\epsilon \leq \sigma_{_0}$ and the alternative hypothesis $H_{a}\,:\, \sigma_\epsilon > \sigma_{_0}\,$, where $\sigma_{_0}$ denotes the specified threshold for acceptable variability.\smallskip

To test the hypothesis $H_{_0}\,:\,\sigma_\epsilon \leq \sigma_{_0}$,the test statistics $SS_\epsilon/\sigma^2_{_0}$ is employed. Under the condition {$\sigma_\epsilon = \sigma_{_0}$}, this statistic has the same distribution as $W = SS_\epsilon/\sigma_\epsilon^2$, which is chi-square distributed with $\mathrm{df}_\epsilon$ degrees of freedom.  Typically, a large observed value of $SS_\epsilon$ provides evidence in favor of the alternative hypothesis  $\sigma_\epsilon > \sigma_{_0}$. The corresponding $p$-value for the test can be computed as 
\begin{align}
    p\textrm{-value} = 1-\Pr\left(W < \mfrac{1}{\sigma^2_0}\,{\sum_{i=1}^a\sum_{j=1}^r (y_{ij} - \overline{y}_{i\cdot})^2}\right). 
\end{align}

In measurement system analysis, it is often necessary to test whether the relative magnitude of the variance components attributed to the units and measurement system exceeds a specified threshold, denoted as $\rho_{_0}$~\citep{donner1987sample,steiner2005statistical}. For this purpose, the null hypothesis is formulated as  $H_{_0}\,:\, \rho \leq \rho_{_0}\,$, with the alternative hypothesis $H_{a}\,:\,\rho >\rho_{_0}\,$.  

For this hypothesis test, the test statistic $(1+r\rho_{_0})^{-1}MS_u/MS_\epsilon$ is employed. Under the null hypothesis $\rho=\rho_{_0}$, this statistic follows an $F$-distribution with degrees of freedom $\mathrm{df}_{u}$ and $\mathrm{df}_\epsilon$. A large observed value of $(1+r\rho_{_0})^{-1}MS_u/MS_\epsilon$ provides evidence in favor of the alternative hypothesis $H_{a}\,:\,\rho >\rho_{_0}\,$.  The $p$-value for this test is computed as
\begin{align}
    p\mbox{-value} = 1-\Pr\left(F <\frac{r\sum_{i=1}^a (y_{i\cdot} - \overline{y}_{\cdot\cdot})^2}{(1+r\rho_{_0})\sum_{i=1}^a\sum_{j=1}^r (y_{ij} - \overline{y}_{i\cdot})^2}\right).
\end{align}

\subsection{Convergence properties}
\label{sec:Convergence_properties}
This section examines the convergence behavior of the estimators for $\sigma^2_u$, $\sigma^2_\epsilon$, and the ratio of these components as the number of measurements $N=ar$ increases in the one-way model~\eqref{eq:one-way-model}. Specifically, we assess whether these estimators are consistent and explore their convergence in distribution.\smallskip

The asymptotic characteristics of the estimators are analyzed in three scenarios, defined by varying the sample size ($a$) and/or the replication number ($r$):
\setlist{nolistsep}
\begin{enumerate}[label=(\textrm{$\mathscr{S}$\arabic*}), leftmargin=3\parindent]
    \item $a \rightarrow \infty$ with $r$ fixed, 
    \item $a\rightarrow \infty$ and $r\rightarrow \infty$,
    \item $r \rightarrow \infty$ with $a$ fixed.
\end{enumerate}\smallskip
We begin our analysis by investigating the asymptotic behavior of the maximum likelihood estimators. This examination serves as a foundation for understanding and interpreting the convergence properties of ANOVA-based estimators.\smallskip

In Theorem~\ref{Theorem:r-fixed}, we provide insight into the convergence behavior for the maximum likelihood estimates of variance components $\sigma^2_u$ and $\sigma^2_\epsilon$ in the one-way random effect model~\eqref{eq:one-way-model} for asymptotic situation $\mathscr{S}1$.

\begin{theorem}\label{Theorem:r-fixed} 
Let $\widehat{\sigma^2_\epsilon\,}$ and $ \widehat{\sigma^2_{u}}$ denote the maximum likelihood estimators. With a fixed number of measurement replications $r$ and as $a\to\infty$, the following convergence properties hold:\smallskip
\setlist{nolistsep}
\begin{enumerate}[label=\textup{{\alph*})},noitemsep]
    \item $\widehat{\sigma^2_\epsilon\,} \xrightarrow[]{\text{a.s}} {\sigma}^2_\epsilon $, provided that  $\mathrm{E}\left[\epsilon_{ij}^2\right]<\infty$\,,\smallskip
    
    \item $\widehat{\sigma^2_{u}} \xrightarrow[]{\text{a.s}} {\sigma}^2_u $, provided that $\mathrm{E}\left[U_{i}^2\right]<\infty$ and $\mathrm{E}\left[\epsilon_{ij}^2\right]<\infty$\,, \smallskip

    \item Within the normal distributional framework of model~\eqref{eq:one-way-model} and under the condition that $\sigma^2_u>0$, 
    
    \begin{align*}
    \sqrt{a\,}\begin{pmatrix}\begin{array}{c}
    \widehat{\sigma^2_\epsilon\,} - {\sigma}^2_\epsilon\\
   \widehat{\sigma^2_{u}} - {\sigma}^2_u
    \end{array}\end{pmatrix}
    \xrightarrow[]{\text{d}} \mathcal{N}\left(\bzero,\left(\begin{array}{cc}
      \sigma_{11} &  \sigma_{12} \\
      \sigma_{21}   & \sigma_{22} \end{array}\right)\right),
    \end{align*}
where the elements of the covariance matrix are 
\begin{gather*}
\sigma_{11} = \mfrac{2\,\sigma^4_\epsilon}{r-1}\,,\quad \sigma_{12}=\mfrac{-2\,\sigma^4_\epsilon}{r(r-1)}\,,\\ 
\mbox{and }\;\sigma_{22}= 2(\sigma^2_u+r^{-1}{\sigma^2_\epsilon})^2 +\mfrac{2\sigma^4_\epsilon}{r^2(r-1)}\,\cdot
\end{gather*}   

\end{enumerate}
\end{theorem}
\begin{proof}
    See Appendix.
\end{proof}

The almost sure convergence properties established in Theorem~\ref{Theorem:r-fixed} rely solely on the assumptions of independence and identical distributions of all $U_i$'s and  $\epsilon_{ij}$'s, as well as the finiteness of their second moments. Notably, these results are independent of the specific distributions of $U_i$'s and $\epsilon_{ij}$'s. Moreover, without assuming the distributions of $U_i$ and $\epsilon_{ij}$'s, part (c) of Theorem~\ref{Theorem:r-fixed} ensures convergence to a normal distribution, provided that $U_i^2$ and $\epsilon_{ij}^2$ have finite variances. However, in such cases, the covariance matrix will differ, as it depends on the underlying distributions of $U_i$'s and $\epsilon_{ij}$'s.\smallskip

In Theorem~\ref{Theorem:ch2-rho}, we examine the convergence behavior of $\widehat{\rho}_{_{\scriptsize\mathrm{MLE}}}$ as the sample size $a$ becomes asymptotically large.

\begin{theorem}\label{Theorem:ch2-rho}
As $a\to \infty$ and for any fixed number of measurement replications $r$, the following convergence properties hold for $\widehat{\rho}_{_{\scriptsize\mathrm{MLE}}}$,
\setlist{nolistsep}
\begin{enumerate}[label=\textup{{\alph*})},noitemsep]
    \item $\widehat{\rho}_{_{\scriptsize\mathrm{MLE}}} \xrightarrow[]{\text{a.s}} \rho $, provided that $\mathrm{E}\left[\epsilon_{ij}^2\right]$ and $\mathrm{E}\left[U_{ij}^2\right]$ are finite,  \vspace{1.5mm}

    \item  $\sqrt{a\,}\left(\widehat{\rho}_{_{\scriptsize\mathrm{MLE}}}-\rho\right) \xrightarrow[]{\text{d}}\mathcal{N}\left(0,\sigma^2_\rho\right)$, where $\sigma^2_\rho = \mfrac{2}{r^2}\left[\left(1+r\rho\right)^2+\mfrac{1}{r-1}\right]$, assuming $\sigma^2_u$ is positive.
\end{enumerate}
\end{theorem}
\begin{proof}
    Applying Slutskey's theorem to parts (a) and (b) of Theorem~\ref{Theorem:r-fixed} establishes the result in part (a). Applying Slutskey's theorem to the marginal asymptotic distribution of the estimator for the variance component $\sigma^2_{u}$, combined with the almost sure convergence of the estimator for $\sigma^2_\epsilon$ to its true value $\sigma^2_\epsilon$, establishes the result in part (b).
\end{proof}

Therefore, the rate of convergence of $\widehat{\rho}_{_{\scriptsize\mathrm{MLE}}}$ is $a^{-1/2}$. Corollary~\ref{cor:a-r-infty}, which examines the convergence of $\widehat{\sigma^2_u}\,$, $\widehat{\sigma^2_\epsilon\,}$ and $\widehat{\rho}_{_{\scriptsize\mathrm{MLE}}}$ as both the sample size $a$ and number of measurement replications $r$ grow large, is derived from the results established in Theorems~\ref{Theorem:r-fixed} and~\ref{Theorem:ch2-rho}.

\begin{corollary}\label{cor:a-r-infty} 
Under the condition that $\sigma^2_u>0$ and the normality of the distribution in model~\eqref{eq:one-way-model}, as $a\to\infty$ and $r\to\infty$ the following convergence properties hold:
\setlist{nolistsep}
\begin{enumerate}[label=\textup{{\alph*})},noitemsep]
    \item The joint distribution of $\big(\widehat{\sigma^2_\epsilon\,},\, \widehat{\sigma^2_u}
\big)^\top$ converges in the following manner, 
    \begin{align*}
     \sqrt{a\,}\begin{pmatrix}
    \sqrt{r\,}\big(\widehat{\sigma^2_\epsilon\,} - {\sigma}^2_\epsilon\big)\\[0.5ex]
   \widehat{\sigma^2_u} - {\sigma}^2_u
    \end{pmatrix}
    \xrightarrow[]{\text{d}} \mathcal{N}\left(\bzero,\begin{pmatrix}
      2\sigma^4_\epsilon &  0 \\
      0   & 2\sigma^4_u\end{pmatrix}\right),
    \end{align*}
    \item $\sqrt{a\,}\left(\widehat{\rho}_{_{\tiny\mathrm{MLE}}}-\rho\right) \xrightarrow[]{\text{d}}\mathcal{N}\left(0,2\rho^2\right) \,$.
\end{enumerate}
\end{corollary}
 Next, we proceed to analyze the asymptotic behavior under the scenario $\mathscr{S}3$. 

\begin{theorem}\label{Theorem:a-fixed}
Given a fixed sample size $a$, as  $r\rightarrow \infty$ the following convergence properties hold:
\setlist{nolistsep}
\begin{enumerate}[label=\textup{{\alph*})},noitemsep]
\item  $\widehat{\sigma^2_\epsilon\,} \xrightarrow[]{\text{a.s}} {\sigma}^2_\epsilon $, provided that  $\mathrm{E}\left[\epsilon_{ij}^2\right]<\infty\,$,\smallskip

\item $\sqrt{ar\,}\left(\widehat{\sigma^2_\epsilon\,} - {\sigma}^2_\epsilon\right)\xrightarrow[]{\text{d}} \mathcal{N} \left(0,\tau_\epsilon^2\right)$, where $\tau_\epsilon^2 = \mathrm{Var}\left[\epsilon_{ij}^2\right]$, provided that $\tau_\epsilon^2 <\infty\,$,\smallskip

\item $\mfrac{a}{\sigma^2_u}\,\widehat{\sigma^2_u}\xrightarrow[]{\text{d}}\chi^2_{a-1}\,$, assuming $\sigma^2_u>0$,\smallskip

\item $\mfrac{a\,}{\rho}\,\widehat{\rho}_{_{\scriptsize\mathrm{MLE}}}\xrightarrow[]{\text{d}}\chi^2_{a-1}\,$, assuming $\sigma^2_u>0\,$.

\end{enumerate}
\end{theorem}

\begin{proof}
    See the appendix.
\end{proof}

Combining the results of Theorems~\ref{Theorem:r-fixed} to~\ref{Theorem:a-fixed}, we conclude that as the number of measurements grows, the probability of obtaining an estimate at the boundary of the parameter space approaches zero. This conclusion is formalized in Theorem~\ref{Theorem:positive-sigma-s}.

\begin{theorem}\label{Theorem:positive-sigma-s}
As the number of measurements $N\to \infty$, and under the conditions specified by model \eqref{eq:one-way-model}, the probabilities of obtaining estimates on the boundaries of the parameter space approach zero for both non-negative ANOVA and maximum likelihood estimation methods.
\end{theorem}

\begin{proof}
    Refer to the appendix.
\end{proof}


\subsection{Confidence intervals}
\label{sec:confidence_intervals}
We investigate the construction of confidence intervals for the variance components $\sigma^2_u$ and $\sigma^2_\epsilon$ as well as the parameter $\rho$, within the context of the measurement system assessment study. Under the assumptions of the one-way model~\eqref{eq:one-way-model}, exact confidence intervals for $\sigma^2_\epsilon$ and $\rho$ are available~\cite{scheffe1959analysis}. However, exact confidence intervals for the variance component $\sigma^2_u\,$ are not straightforward to derive. To address this, we illustrate the procedure for constructing approximate confidence intervals using results from large-sample theory.

\subsubsection*{A.~Exact confidence intervals}
Under the normality assumption of $\epsilon_{ij}$ in the one-way random effect model of~\eqref{eq:one-way-model}, the statistic ${SS_\epsilon}/{\sigma^2_\epsilon}$ follows a $\chi^2_{a(r-1)}$ distribution. This property facilitates the derivation of the exact $(1-\alpha)100\%$ confidence interval for ${\sigma}^2_\epsilon$ as shown below, 
\begin{align}\label{eq:CI-e}
\mathcal{CI}_{\sigma^2_\epsilon}=
\left[\mfrac{SS_\epsilon}{\chi^2_{a(r-1),1-{\alpha}/{2}}} ,\, \mfrac{SS_\epsilon}{\chi^2_{a(r-1),\,{\alpha}/{2}} }\right],
\end{align}
where $\chi^2_{\nu,1-{\alpha}/{2}}$ and $\chi^2_{\nu,\,{\alpha}/{2}}$ are, respectively, the upper and lower critical values of a chi-squared distribution with $\nu$ degrees of freedom at the $\frac{\alpha}{2}$--level.

Building on this confidence interval for ${\sigma}^2_\epsilon$, the exact confidence interval for the parameter PTR, as defined in \eqref{eq:PTR} is given by
\begin{align}\label{eq:CI-PTR}
\mathcal{CI}_{\text{PTR}}=
\left[\mfrac{\kappa}{UL-LL}\left(\mfrac{SS_\epsilon}{\chi^2_{a(r-1),1-{\alpha}/{2}}}\right)^{1/2} ,\, \mfrac{\kappa}{UL-LL}\left(\mfrac{SS_\epsilon}{\chi^2_{a(r-1),\,{\alpha}/{2}} }\right)^{1/2}\right],
\end{align}

Under the assumptions of normality and independence of $U_i$'s and $\epsilon_{ij}$'s, the exact confidence interval for the ratio of the variance components $\sigma^2_u$ and $\sigma^2_\epsilon$ can be derived. From~\eqref{eq:F}, it follows that ${(1+r\rho)^{-1}}\mfrac{MS_u}{MS_\epsilon}$ follows an $F$-distribution with $a-1$ and $a(r-1)$ degrees of freedom. Therefore, the exact $(1-\alpha)100\%$ confidence interval for $\rho$ is given by,
\begin{align}\label{eq:CI-rho}
    \mathcal{CI}_{\rho}=\left[\mfrac{1}{r}\Big(\mfrac{MS_u/MS_\epsilon}{F_{a-1,\,a(r-1),\,1-{\alpha}/2}}-1 \Big),\,  \mfrac{1}{r}\Big(\mfrac{MS_u/MS_\epsilon}{F_{a-1,\,a(r-1),\,{\alpha}/2}}-1 \Big)\right].
\end{align}
Here, $F_{a-1,\,a(r-1),\,1-{\alpha}/2}$ and $F_{a-1,\,a(r-1),\,{\alpha}/2}$ denote, the upper and lower critical values of an $F$-distribution with $a-1$ and $a(r-1)$ degrees of freedom at the $\frac{\alpha}{2}$-level, respectively. 

Building on this confidence interval for $\rho$, the exact confidence intervals for the parameters R\&R ratio, SNR, and ICC, are expressed as follows:\smallskip
\begin{itemize}
    \item {R\&R ratio confidence interval:} The R\&R ratio is given by \mbox{$\text{R\&R} = (1+\rho)^{-1/2}$}. Using the bounds for $\rho$ from~\eqref{eq:CI-rho}, the exact confidence interval for $\text{R\&R}$ is, 
    \begin{align*}\label{eq:CI-R&R}
        \mathcal{CI}_{\text{R\&R}}=\left[\Big(1+\mfrac{MS_u/MS_\epsilon}{rF_{a-1,\,a(r-1),\,{\alpha}/2}}-\mfrac{1}{r} \Big)^{-1/2},\,  \Big(1+\mfrac{MS_u/MS_\epsilon}{rF_{a-1,\,a(r-1),\,1-{\alpha}/2}}-\mfrac{1}{r} \Big)^{-1/2}\right].
    \end{align*}
    \item {SNR confidence interval:} The SNR is given by $\rho^{1/2}$. Using the confidence interval for $\rho$, the corresponding confidence interval for SNR is,
    \begin{align*}
        \mathcal{CI}_{\text{SNR}}=\left[\mfrac{1}{\sqrt{r}}\Big(\mfrac{MS_u/MS_\epsilon}{F_{a-1,\,a(r-1),\,1-{\alpha}/2}}-1 \Big)^{1/2},\,  \mfrac{1}{\sqrt{r}}\Big(\mfrac{MS_u/MS_\epsilon}{F_{a-1,\,a(r-1),\,{\alpha}/2}}-1 \Big)^{1/2}\right].
    \end{align*}
    \item {ICC confidence interval:} The ICC is defined as $\text{ICC}=1/(1+\rho^{-1})$ . The exact confidence interval for ICC is, 
    \begin{align*}
\mathcal{CI}_{\text{ICC}}=\left[\mfrac{1}{1+r\Big(\mfrac{MS_u/MS_\epsilon}{F_{a-1,\,a(r-1)},\,1-{\alpha}/2}-1\Big)^{-1}}\,,\,  \mfrac{1}{1+r\Big(\mfrac{MS_u/MS_\epsilon}{F_{a-1,\,a(r-1),\,{\alpha}/2}}-1 \Big)^{-1}}\right].
    \end{align*}
\end{itemize}

\subsubsection*{B.~Approximate confidence intervals}
An exact confidence interval for $\sigma^2_u$ is not available. However, large sample intervals can be constructed as suggested by~\cite{montgomery1993gaugeII,borror1997confidence}.

Leveraging the result of Theorem~\ref{Theorem:r-fixed}, a Wald-type confidence interval for $\sigma^2_u$ can be established. The statistic $\sqrt{a\,}(\widehat{\sigma^2_{u}} - {\sigma}^2_u)/\sqrt{\widehat{\sigma_{22}}}$ is asymptotically pivotal, yielding the following Wald-type confidence interval for $\sigma^2_u$ at $1-\alpha$ level,
\begin{align}
    \left[\widehat{\sigma^2_{u}}-\mfrac{z_{1-\alpha/2}\sqrt{\widehat{\sigma_{22}}}}{\sqrt{a\,}},\widehat{\sigma^2_{u}}+\mfrac{z_{1-\alpha/2}\sqrt{\widehat{\sigma_{22}}}}{\sqrt{a\,}}\right].
\end{align}
However, this confidence interval of $\sigma^2_u$ is not range-preserving, as its lower bound can occasionally be negative. To address this limitation, one can redefine any negative bounds to be zero, resulting in the interval 
\begin{align}
    \mathcal{CI}_{\sigma^2_u,\mathrm{Wald}}=\left[\max\left(0,\widehat{\sigma^2_{u}}-\mfrac{z_{1-\alpha/2}\sqrt{\widehat{\sigma_{22}}}}{\sqrt{a\,}}\right),\widehat{\sigma^2_{u}}+\mfrac{z_{1-\alpha/2}\sqrt{\widehat{\sigma_{22}}}}{\sqrt{a\,}}\right].
\end{align}
To ensure a range-preserving confidence interval for $\sigma^2_u\,$, a Wald confidence interval can be constructed based on the logarithmic transformation of $\widehat{\sigma^2_u}\,$. The limits are then inverted to approximate a confidence interval for $\sigma^2_u$. Applying the delta method to the asymptotic distribution derived in Theorem~\ref{Theorem:r-fixed} yields the following asymptotic convergence result,
\begin{align}
    \sqrt{a}\left\{\log\big(\widehat{\sigma^2_u\,}\big) - \log\big({\sigma}^2_u\big)\right\}\xrightarrow[]{\text{d}} \mathcal{N} \left(0,\mfrac{\sigma_{22}}{\sigma^4_u}\right),
\end{align}
as $a\to \infty$. Consequently the quantity $\sqrt{a\,}\big\{\log(\widehat{\sigma^2_{u}}) - \log({\sigma}^2_u)\big\}\widehat{\sigma^2_{u}}/\sqrt{\widehat{\sigma_{22\,}}}$ is asymptotically pivotal, provided that $\widehat{\sigma^2_{u}}>0$. This result leads to the construction of the following $(1-\alpha)100\%$ confidence interval for $\sigma^2_u$,
\begin{align}\label{eq:CI-log-transform}
    \mathcal{CI}_{\sigma^2_u,\,\log}= \left[\exp\left\{\log(\widehat{\sigma^2_{u}})-\mfrac{z_{1-{\alpha}/{2}} \sqrt{\widehat{\sigma_{22}}}}{\widehat{\sigma^2_{u}}\sqrt{a\,}}\right\}, \exp\left\{\log(\widehat{\sigma^2_{u}})+\mfrac{z_{1-{\alpha}/{2}} \sqrt{\widehat{\sigma_{22}}}}{\widehat{\sigma^2_{u}}\sqrt{a\,}}\right\}\right].
\end{align}
It is worth noting that when the number of measurements or the signal-to-noise ratio is small, the log-transformed confidence interval for $\sigma^2_u$ may not offer an accurate approximation. This limitation arises, in part, from the increased likelihood of obtaining an estimate $\widehat{\sigma^2_u}$ that is close to zero. As demonstrated by~\cite{ebadi2023phase}, applying the Cornish–Fisher expansion to adjust the percentiles can lead to improved results. Exploring these adjustments in greater detail represents a promising direction for future research.

We can derive a large sample confidence interval for $\sigma^2_u$, using the result of Theorem~\ref{Theorem:a-fixed}. This asymptotic confidence interval provides a $(1-\alpha)100\%$ coverage probability for $\sigma^2_u$ and is expressed as 
\begin{align}
    \mathcal{CI}_{\sigma^2_u,\,\mathrm{chi}}=\left[\mfrac{a\,\widehat{\sigma^2_u}}{\chi^2_{a-1,\,1-\alpha/2}},\,\mfrac{a\,\widehat{\sigma^2_u}}{\chi^2_{a-1,\,\alpha/2}}\right].
\end{align}

Table~\ref{table:CI-sigma2_u} summarizes the bounds of the three approximate confidence intervals for $\sigma^2_u$ discussed in this section.

\vspace{2mm}
\begin{table}[htb!]
    \centering
    \caption{Bounds of approximate confidence intervals on $\sigma^2_u$ in the one-way random effect model, derived using asymptotic distributions.}
    \vspace{1mm}
    \scalebox{0.85}{
    \begin{tblr}{
    colspec ={c c c},
    hline{1,5} = {},
    vline{1,2,4}={},
    vline{3} ={dotted},
    cell{1}{1-3} = {blue!10},
    }
     {Interval} & {Lower limit}  & {Upper limit} \\  \hline\hline
    {$\mathcal{CI}_{\sigma^2_u,\mathrm{Wald}}$} & $\max\left(0,\widehat{\sigma^2_{u}}-\mfrac{z_{1-\alpha/2}\sqrt{\widehat{\sigma_{22}}}}{\sqrt{a\,}}\right)$ & $\widehat{\sigma^2_{u}}+\mfrac{z_{1-\alpha/2}\sqrt{\widehat{\sigma_{22}}}}{\sqrt{a\,}}$ \\
    [1ex]
    {$\mathcal{CI}_{\sigma^2_u,\,\log}$} &  $\exp\left\{\log(\widehat{\sigma^2_{u}})-\mfrac{z_{1-{\alpha}/{2}} \sqrt{\widehat{\sigma_{22}}}}{\widehat{\sigma^2_{u}}\sqrt{a\,}}\right\}$ & $\exp\left\{\log(\widehat{\sigma^2_{u}})+\mfrac{z_{1-{\alpha}/{2}} \sqrt{\widehat{\sigma_{22}}}}{\widehat{\sigma^2_{u}}\sqrt{a\,}}\right\}$ \\
    [1ex]
    {$\mathcal{CI}_{\sigma^2_u,\mathrm{chi}}$} & $\mfrac{a\,\widehat{\sigma^2_u}}{\chi^2_{a-1,\,1-\alpha/2}}$ & $\mfrac{a\,\widehat{\sigma^2_u}}{\chi^2_{a-1,\,\alpha/2}}$  \\
     \end{tblr}
 }
 \label{table:CI-sigma2_u}
\end{table}

\subsubsection*{C.~Comparison of confidence intervals}
We conducted Monte Carlo simulation studies to assess the performance of the three confidence intervals for $\sigma^2_u$  listed in Table~\ref{table:CI-sigma2_u}. The simulations considered six combinations of $a$ and $r$, maintaining a constant total of $N=96$ measurements. For each $(a,r)$ combination, we examined three variance settings: $(\sigma^2_u, \sigma^2_\epsilon) = (0.5, 1)$, $(0.5, 0.5)$, and $(0.5, 0.1)$, representing low, moderate, and high signal-to-noise ratio scenarios, respectively.

In each scenario, $5\times 10^5$ datasets were simulated, with $U_i$ and $\epsilon_{ij}$ drawn independently from normal distributions with variances $\sigma^2_u$ and $\sigma^2_\epsilon$. These simulations evaluated the confidence intervals' performance across different signal-to-noise ratio levels and varying $(a, r)$ combinations.

We evaluated the coverage probability of each of the three confidence intervals by determining how often it successfully included the true value of $\sigma^2_u$. For the log-transformed confidence interval, we focused on samples where $\widehat{\sigma^2_u} > 0.01$, as this method is sensitive to estimates close to zero. To assess the likelihood of encountering such cases, Table~\ref{table:Pr_sigma_u} presents the probabilities $\Pr(\widehat{\sigma^2_u} < 0.01)$ for the scenarios considered in the study.

\begin{table}[htb!]
\centering
\caption{$\Pr(\,\widehat{\sigma^2_u}<0.01)$ with $N=96$ measurements.}
\vspace{1mm}
\scalebox{0.8}{
\begin{tabular}
{l  llllll}
 \hline
  &\multicolumn{6}{c}{$(a,r)$} \\ 
  \cmidrule(rl){2-7}
    $(\sigma^2_u,\sigma^2_\epsilon)$ & $(6,16)$ & $(8,12)$ & $(12,8)$ & $(24,4)$ & $(32,3)$ & $(48,2)$ \\ 
 \hline [0.4ex] 
 \hline
         $(0.5,1)$ & $2.15 \times 10^{-2}$ &	$1.18 \times 10^{-2}$ &	$5.36 \times 10^{-3}$ & $3.72 \times 10^{-3}$ & $5.27 \times 10^{-3}$ & $1.25 \times 10^{-2}$ \\ 
         $(0.5,0.5)$ & $6.55\times 10^{-3}$ &	$2.26\times 10^{-3}$ &	$4.20\times 10^{-4}$ &	$4.0\times 10^{-05}$ & $8.0\times 10^{-5}$ &	$1.90\times 10^{-4}$  \\ 
         $(0.5,0.1)$ & $7.90\times 10^{-4}$ & $9.0\times 10^{-5}$ & 0 &	0 & 0	& 0 \\ [0.5ex]
 \hline
 \hline
 \end{tabular}
 }
\label{table:Pr_sigma_u}
\end{table}


Tables~\ref{table:ACI-alpha-0.1} and~\ref{table:ACI-alpha-0.05} present the results of our numerical analysis, focusing on the $90\%$ and $95\%$ asymptotic confidence intervals and the average lengths of these intervals, respectively. The highest simulated coverage probabilities among the three intervals are highlighted for clarity.

Our findings reveal that the log-transformed confidence intervals, $\mathcal{CI}_{\sigma^2_u,\,\log}$, generally offer superior coverage compared to the Wald-type intervals. However, it is important to note that the reliability of $\mathcal{CI}_{\sigma^2_u,\,\log}$ diminishes under conditions of low signal-to-noise ratio or a small number of measurement replications, often resulting in significantly increased interval lengths.

In moderate to high signal-to-noise ratio settings, $\mathcal{CI}_{\sigma^2_u,\,\log}$ consistently outperforms $\mathcal{CI}_{\sigma^2_u,\,\mathrm{chi}}$ in achieving the target coverage level, particularly for large sample sizes. Conversely, in scenarios with a greater number of replications, $\mathcal{CI}_{\sigma^2_u,\,\mathrm{chi}}$ demonstrates superior performance, in terms of the coverage probability.

\section{Conclusion}\label{sec:Conclusion}
This paper presents a comprehensive theoretical investigation of the statistical methods used in measurement system assessment, focusing on the univariate one-way random effects model. We analyze several estimators for variance components and key parameters in measurement system assessment.

The ANOVA-based estimator, while widely used, inherently has the potential to produce negative estimates—a limitation that becomes less pronounced with increasing SNR or the number of measurements $N=ar$. Our analysis also examines the bias and sampling variance properties of the estimators for variance components and the SNR. Due to the complexity of the underlying equations, deriving the bias and variance analytically for the maximum likelihood estimators is infeasible. However, the ANOVA-based SNR estimator consistently exhibits a positive bias, which highlights a key characteristic of its performance. Numerical studies revealed that the maximum likelihood estimator of the SNR maintained an approximately zero bias when the replication number $r$ equals $3$, regardless of the sample size--a desirable feature for reliable measurement system assessments. Moreover, a comparative analysis demonstrated that  $\widehat{\rho}_{\ds{\tiny\mbox{MLE}}}$ consistently exhibited lower SE compared to $\widehat{\rho}_{\ds{\tiny\mbox{NANOVA}}}$ across study plans with a total of $60$ measurements. Notably, the smallest relative standard error for $\widehat{\rho}_{\ds{\tiny\mbox{MLE}}}$ was observed under a study plan with a sample size of $20$ and $3$ measurement replications.

We investigated the asymptotic behavior of the maximum likelihood estimators across three scenarios, defined by variations in the sample size and/or the replication number. To assess the statistical significance of the point estimates and support informed decision-making regarding the measurement system, we explored procedures for conducting specific hypothesis tests and constructing confidence intervals. 

The one-way model serves as a simple yet powerful framework for many practical applications. The insights gained from this work establish a theoretical foundation for expanding the investigation to more complex models, multivariate data, functional data, and advanced experimental designs.

\begin{table}[!htb]
\caption{Simulated $90\%$ confidence intervals for $\sigma^2_u$ with $\alpha = 0.1$ and $N=96$ measurements.}
 \centering
\scalebox{0.75}{
\begin{tblr}{
colspec={c c c c c c c},
cell{1}{1} = {r = 2 , c = 7}{halign = c, valign = b},
cell{3}{2} = {c = 3}{halign = c,gray!10},
cell{3}{5} = {c = 3}{halign = c,gray!10},
cell{5-10}{3} = {blue!15},
cell{4}{2-7} = {lightgray},
hline{3,5} ={},
vline{2,5} ={dotted}
}
  (a)~$\sigma^2_u = 0.5,\, \sigma^2_\epsilon = 1$ & & & & & & \\
  & & & & &  &\\
  & Coverage probability & & & Mean interval width & & &  \\ 
  $(a,r)$ & $\mathcal{CI}_{\sigma^2_u,\mathrm{Wald}}$ & $\mathcal{CI}_{\sigma^2_u,\log}$ & $\mathcal{CI}_{\sigma^2_u,\mathrm{chi}}$ & $\mathcal{CI}_{\sigma^2_u,\mathrm{Wald}}$ & $\mathcal{CI}_{\sigma^2_u,\log}$ & $\mathcal{CI}_{\sigma^2_u,\mathrm{chi}}$ \\
  $(6,16)$ & 0.687 &	0.903 &	0.828 & 0.851 &	1.148 &	1.906 \\ 
$(8,12)$ & 0.735 & 0.919 & 0.818 & 0.825 & 1.063 & 1.334  \\ 
$(12,8)$ & 0.786 & 0.941 &	0.790 &	0.765 &	1.131 &	0.902 \\ 
$(24,4)$ & 0.846 & 0.962 & 0.701 & 0.690 &	$1.802\times 10^{2}$ &	0.540 \\ 
$(32,3)$ & 0.866 & 0.959 & 0.646 & 0.682 &	$6.832\times 10^{4}$ &	0.450 \\ 
$(48,2)$ & 0.891 & 0.943 & 0.533 & 0.713 & 	$1.599\times 10^{11}$	& 0.354 \\ [0.5ex]
 \hline 
 \end{tblr}
 }
\centering
\scalebox{0.8}{
\begin{tblr}{
colspec={c c c c c c c},
cell{1}{1} = {r = 2 , c = 7}{halign = c, valign = b},
cell{3}{2} = {c = 3}{halign = c,gray!10},
cell{3}{5} = {c = 3}{halign = c,gray!10},
cell{5-6}{4} = {blue!15},
cell{7-10}{3} = {blue!15},
cell{4}{2-7} = {lightgray},
hline{3,5} ={},
vline{2,5} ={dotted}
}
    (b)~$\sigma^2_u = 0.5,\, \sigma^2_\epsilon = 0.5$ & & & & &  &\\
  & & & & & & \\
  & Coverage probability & & & Mean interval width & & &  \\ 
  $(a,r)$ & $\mathcal{CI}_{\sigma^2_u,\mathrm{Wald}}$ & $\mathcal{CI}_{\sigma^2_u,\log}$ & $\mathcal{CI}_{\sigma^2_u,\mathrm{chi}}$ & $\mathcal{CI}_{\sigma^2_u,\mathrm{Wald}}$ & $\mathcal{CI}_{\sigma^2_u,\log}$ & $\mathcal{CI}_{\sigma^2_u,\mathrm{chi}}$ \\
    $(6,16)$ & 0.687 &	0.822 &	 0.864 & 0.828 & 1.004 & 1.929 \\ 
    $(8,12)$ & 0.735 &	0.847 &	0.859 &	0.778 & 0.895 &	1.350  \\ 
    $(12,8)$ & 0.785 & 0.872 & 0.845 &	0.693 & 0.765	& 0.912 \\ 
    $(24,4)$ & 0.842 & 0.901 & 0.799 & 0.573 & 0.612 &	0.546 \\ 
    $(32,3)$ & 0.859 &	0.913 & 0.766 & 0.541 &	0.577 &	0.454 \\ 
    $(48,2)$ & 0.881 &	0.939 & 0.695 &	0.522 &	3.633 &	0.358 \\ [0.5ex]
 \hline 
 \end{tblr}
 }
\centering
\scalebox{0.8}{
\begin{tblr}{
colspec={c c c c c c c},
cell{1}{1} = {r = 2 , c = 7}{halign = c, valign = b},
cell{3}{2} = {c = 3}{halign = c,gray!10},
cell{3}{5} = {c = 3}{halign = c,gray!10},
cell{5-8}{4} = {blue!15},
cell{9-10}{3} = {blue!15},
cell{4}{2-7} = {lightgray},
hline{3,5} ={},
vline{2,5} ={dotted}
}
    (c)~$\sigma^2_u = 0.5,\, \sigma^2_\epsilon = 0.1$ & & & & &  &\\
  & & & & & & \\
  & Coverage probability & & & Mean interval width & & &  \\ 
  $(a,r)$ & $\mathcal{CI}_{\sigma^2_u,\mathrm{Wald}}$ & $\mathcal{CI}_{\sigma^2_u,\log}$ & $\mathcal{CI}_{\sigma^2_u,\mathrm{chi}}$ & $\mathcal{CI}_{\sigma^2_u,\mathrm{Wald}}$ & $\mathcal{CI}_{\sigma^2_u,\log}$ & $\mathcal{CI}_{\sigma^2_u,\mathrm{chi}}$ \\
    $(6,16)$ & 0.687 &	0.798 &	0.893 &  0.800 &	0.931 &	1.949 \\ 
    $(8,12)$ & 0.735 &	0.827 &	0.892 &  0.732 &	0.821 &	1.363 \\ 
    $(12,8)$ & 0.785 &	0.852 &	0.889 &  0.631 &  0.682 &	0.921 \\ 
    $(24,4)$ & 0.840 &  0.878 & 0.880	& 0.478	& 0.498	& 0.550 \\ 
    $(32,3)$ & 0.855 &	0.885 & 0.875 &  0.425 &  0.439 & 0.458\\ 
    $(48,2)$ & 0.871 &	0.893 & 0.863 &  0.364 & 0.372 & 0.361 \\ [0.5ex]
 \hline 
 \end{tblr}
 }
\label{table:ACI-alpha-0.1}
\end{table}

\begin{table}[!htb]
\centering
\caption{Simulated $95\%$ confidence intervals for $\sigma^2_u$ with $N=96$ measurements.}
\scalebox{0.75}{
\begin{tblr}{
colspec={c c c c c c c},
cell{1}{1} = {r = 2 , c = 7}{halign = c, valign = b},
cell{3}{2} = {c = 3}{halign = c,gray!10},
cell{3}{5} = {c = 3}{halign = c,gray!10},
cell{5-10}{3} = {blue!15},
cell{4}{2-7} = {lightgray},
hline{3,5} ={},
vline{2,5} ={dotted}
}
  (a)~$\sigma^2_u = 0.5,\, \sigma^2_\epsilon = 1$ & & & & & & \\
  & & & & &  &\\
  & Coverage probability & & & Mean interval width & & &  \\ 
  $(a,r)$ & $\mathcal{CI}_{\sigma^2_u,\mathrm{Wald}}$ & $\mathcal{CI}_{\sigma^2_u,\log}$ & $\mathcal{CI}_{\sigma^2_u,\mathrm{chi}}$ & $\mathcal{CI}_{\sigma^2_u,\mathrm{Wald}}$ & $\mathcal{CI}_{\sigma^2_u,\log}$ & $\mathcal{CI}_{\sigma^2_u,\mathrm{chi}}$ \\
    $(6,16)$ & 0.728 & 0.992 & 0.889 & 0.936 & 1.566 & 2.740 \\ 
    $(8,12)$ & 0.776 & 0.991 &	0.882 &	0.929 &	1.570 &	1.809 \\ 
    $(12,8)$ & 0.827 & 0.991 &	0.860 &	0.893 &	3.248 &	1.164 \\ 
    $(24,4)$ & 0.889 & 0.984 &	0.781 &	0.813 &	$9.957 \times 10^3$	& 0.667 \\ 
    $(32,3)$ & 0.910 & 0.980 &	0.726 &	0.802 & $	1.246\times 10^7$ &	0.551 \\ 
    $(48,2)$ & 0.940 & 0.967 &	0.611 &	0.833 &	$6.611\times 10^{14}$ & 0.430 \\ [0.5ex]
 \hline 
 \end{tblr}
 }
\centering
\scalebox{0.8}{
\begin{tblr}{
colspec={c c c c c c c},
cell{1}{1} = {r = 2 , c = 7}{halign = c, valign = b},
cell{3}{2} = {c = 3}{halign = c,gray!10},
cell{3}{5} = {c = 3}{halign = c,gray!10},
cell{5-6}{4} = {blue!15},
cell{7-10}{3} = {blue!15},
cell{4}{2-7} = {lightgray},
hline{3,5} ={},
vline{2,5} ={dotted}
}
    (b)~$\sigma^2_u = 0.5,\, \sigma^2_\epsilon = 0.5$ & & & & &  &\\
  & & & & & & \\
  & Coverage probability & & & Mean interval width & & &  \\ 
  $(a,r)$ & $\mathcal{CI}_{\sigma^2_u,\mathrm{Wald}}$ & $\mathcal{CI}_{\sigma^2_u,\log}$ & $\mathcal{CI}_{\sigma^2_u,\mathrm{chi}}$ & $\mathcal{CI}_{\sigma^2_u,\mathrm{Wald}}$ & $\mathcal{CI}_{\sigma^2_u,\log}$ & $\mathcal{CI}_{\sigma^2_u,\mathrm{chi}}$ \\
    $(6,16)$ & 0.728	& 0.895	& 0.921	& 0.911 & 1.282 &	2.773 \\ 
    $(8,12)$ & 0.776 & 0.913 & 0.918 & 0.897 & 1.126 &	1.831  \\ 
    $(12,8)$ & 0.826 &	0.934 & 0.908 &	0.823 & 0.950 & 1.177 \\ 
    $(24,4)$ & 0.885 &	0.959 & 0.870 &	0.683 &	0.756 &	0.675 \\ 
    $(32,3)$ & 0.903 & 	0.969 & 0.842 & 0.644 &	0.834 &	0.556 \\ 
    $(48,2)$ & 0.927 & 	0.977 & 0.776 &	0.621 &	$8.364\times 10^{1}$ &	0.433 \\ [0.5ex]
 \hline 
 \end{tblr}
 }
\centering
\scalebox{0.8}{
\begin{tblr}{
colspec={c c c c c c c},
cell{1}{1} = {r = 2 , c = 7}{halign = c, valign = b},
cell{3}{2} = {c = 3}{halign = c,gray!10},
cell{3}{5} = {c = 3}{halign = c,gray!10},
cell{5-8}{4} = {blue!15},
cell{9-10}{3} = {blue!15},
cell{4}{2-7} = {lightgray},
hline{3,5} ={},
vline{2,5} ={dotted}
}
    (c)~$\sigma^2_u = 0.5,\, \sigma^2_\epsilon = 0.1$ & & & & &  &\\
  & & & & & & \\
  & Coverage probability & & & Mean interval width & & &  \\ 
  $(a,r)$ & $\mathcal{CI}_{\sigma^2_u,\mathrm{Wald}}$ & $\mathcal{CI}_{\sigma^2_u,\log}$ & $\mathcal{CI}_{\sigma^2_u,\mathrm{chi}}$ & $\mathcal{CI}_{\sigma^2_u,\mathrm{Wald}}$ & $\mathcal{CI}_{\sigma^2_u,\log}$ & $\mathcal{CI}_{\sigma^2_u,\mathrm{chi}}$ \\
    $(6,16)$ & 0.728 &	0.861 &	0.944 &	0.892 & 1.178 &	2.802 \\ 
    $(8,12)$ & 0.776 &	0.886 &	0.944 &	0.870 &	1.024 &	1.848 \\ 
    $(12,8)$ & 0.826 &	0.911 &	0.942 &	0.752 &	0.840 &	1.188 \\ 
    $(24,4)$ & 0.882 &	0.932 &	0.936 &	0.570 &	0.604 &	0.680 \\ 
    $(32,3)$ & 0.898 &	0.938 & 0.932 &	0.507 &	0.530 &	0.561 \\ 
    $(48,2)$ & 0.916 & 	0.945 & 0.923 &	0.433 &	0.447 &	0.437 \\ [0.5ex]
 \hline 
 \end{tblr}
 }
\label{table:ACI-alpha-0.05}
\end{table}

\clearpage
\bibliographystyle{tfcad}
\bibliography{uw-ethesis}


\appendix
\section{Proof of convergence properties}\label{Appendix}
We proceed to establish the convergence properties of the estimators introduced in Section~\ref{sec:statistical}.

\subsubsection*{Proof of Theorem~\ref{Theorem:r-fixed}}
(a) Recall that $\widehat{\sigma^2_\epsilon\,} = MS_\epsilon = \mfrac{1}{a(r-1)}\sum_{i=1}^a\sum_{j=1}^r (Y_{ij} - \overline{Y}_{i\cdot})^2$. The random variable $Y_{ij} - \overline{Y}_{i\cdot}$ can be  expressed as $\epsilon_{ij} - \overline{\epsilon}_{i\cdot}$ where $\overline{\epsilon}_{i\cdot}= \frac{1}{r}\sum_{j=1}^r \epsilon_{ij}$. Assuming that the  $\epsilon_{ij}$'s are independent random variables with zero mean and finite second moment $\sigma^2_\epsilon$, it can be shown that the second moment for the random variable $\epsilon_{ij} - \overline{\epsilon}_{i\cdot}$   is  $\mathrm{E}\left[(\epsilon_{ij} - \overline{\epsilon}_{i\cdot})^2\right] = \mfrac{r-1}{r}\sigma^2_\epsilon$. 
Since all $\epsilon_{ij}$'s are identically distributed too, random variables  $\sum_{j=1}^r (\epsilon_{1j} - \overline{\epsilon}_{1\cdot})^2$, $\sum_{j=1}^r (\epsilon_{2j} - \overline{\epsilon}_{2\cdot})^2,\ldots$ are independent, identically distributed, and have a finite mean, 
\begin{align}
    \mathrm{E}\Big[\sum_{j=1}^r (\epsilon_{ij} - \overline{\epsilon}_{i\cdot})^2\Big] = (r-1)\sigma^2_\epsilon,\quad  \mbox{for }\; i = 1,\ldots, a. 
\end{align}
By the {strong law of large numbers}, we have,
\begin{align}
    \mfrac{1}{a}\sum_{i=1}^a \sum_{j=1}^r(\epsilon_{ij} - \overline{\epsilon}_{i\cdot})^2 \xrightarrow[]{\text{a.s}} (r-1)\sigma^2_\epsilon,
\end{align} 
as $a \rightarrow\infty$, for any fixed $r$. This completes the proof for part (a). \smallskip

(b) The maximum likelihood estimate of $\sigma^2_u$ is given by $\widehat{\sigma^2_{u}} = \mfrac{1}{r}\big(\beta^{-1}MS_u-MS_\epsilon\big)$, which can also be expressed as,
\begin{align*}
    \widehat{\sigma^2_{u}} 
    &= \mfrac{1}{a}\sum_{i=1}^a (\overline{Y}_{i\cdot} - \overline{Y}_{\cdot\cdot})^2 - \mfrac{1}{ar(r-1)}\sum_{i=1}^a \sum_{j=1}^r(Y_{ij} - \overline{Y}_{i\cdot})^2. 
\end{align*}
The random variable $\overline{Y}_{i\cdot}$ can be expressed as $\mu + U_i + \overline{\epsilon}_{i\cdot}$. Assuming that all $U_i$'s and all $\epsilon_{ij}$'s are i.i.d., it follows that the random variables $\overline{Y}_{1\cdot}$, $\overline{Y}_{2\cdot},\ldots$ are identically distributed and independent with mean $\mu$ and the variance of $\sigma^2_u + r^{-1}\sigma^2_\epsilon$. Using the consistency of the second sample moment for $\overline{Y}_{1\cdot}$, $\overline{Y}_{2\cdot},\ldots$ we have, 
\begin{align}
    \mfrac{1}{a}\sum_{i=1}^a (\overline{Y}_{i\cdot} - \overline{Y}_{\cdot\cdot})^2  \xrightarrow[]{\text{a.s}} \sigma^2_u+ \mfrac{\sigma^2_\epsilon}{r},
\end{align}
as $a\to \infty$. 
From part (a), we also have that $MS_\epsilon \to \sigma^2_\epsilon$ almost surely as $a\to\infty$. Combining these results, we conclude
\begin{align}
    \widehat{\sigma^2_{u}} \xrightarrow[]{\text{a.s}} {\sigma}^2_u >0,\;\;\textrm{ as  } a\rightarrow \infty ,
\end{align}
which means that the estimator $\widehat{\sigma^2_u}$ converges almost surely to the true value of ${\sigma}^2_u$ as $a\rightarrow \infty$, value of $r$.\smallskip

(c) To prove this result, we begin with the equation,
{\small{
    \begin{align} \label{eq:vec-sigma-e-U} 
    \sqrt{a\,}\left(\!\!\begin{array}{c}
    \widehat{\sigma^2_\epsilon} - {\sigma}^2_\epsilon\\
    \widehat{\sigma^2_{u}} - {\sigma}^2_u
    \end{array}\!\!\right) &= \sqrt{a\,} \left(\!\!\begin{array}{c}
    \mfrac{1}{a(r-1)}\sum_{i=1}^a\sum_{j=1}^r (Y_{ij} - \overline{Y}_{i\cdot})^2 - \sigma^2_\epsilon \\
    \mfrac{1}{a}\sum_{i=1}^a (\overline{Y}_{i\cdot} - \overline{Y}_{\cdot\cdot})^2 - \mfrac{1}{ar(r-1)}\sum_{i=1}^a\sum_{j=1}^r (Y_{ij} - \overline{Y}_{i\cdot})^2 - \sigma^2_u
    \end{array}\!\!\right).
    \end{align}
}}
This equation can be expressed as the sum of two vectors,
{\small{
\begin{align}\label{eq:two-vec}
    \frac{1}{\sqrt{a}}\sum_{i=1}^a\left(\!\!\begin{array}{c} 
    \mfrac{1}{(r-1)}\sum_{j=1}^r (Y_{ij} - \overline{Y}_{i\cdot})^2 - \sigma^2_\epsilon\\
    \mfrac{-1}{r(r-1)}\sum_{j=1}^r (Y_{ij} - \overline{Y}_{i\cdot})^2 + \frac{\sigma^2_\epsilon}{r}
    \end{array}\!\!\right) \;\;\mbox{and}\;\;
 \frac{1}{\sqrt{a}}\sum_{i=1}^a\left(\!\!\begin{array}{c} 
    0 \\
    \left(\overline{Y}_{i\cdot} - \overline{Y}_{\cdot\cdot}\right)^2 - \sigma^2_u -\mfrac{\sigma^2_\epsilon}{r}
    \end{array}\!\!\right).
\end{align}
}}
Using the multivariate central limit theorem, the first term satisfies,
\begin{align}\label{eq:vec1}
    \frac{1}{\sqrt{a}}\sum_{i=1}^a\left(\!\!\begin{array}{c} 
    \frac{1}{r-1}\sum_{j=1}^r (Y_{ij} - \overline{Y}_{i\cdot})^2 - \sigma^2_\epsilon\\
    \frac{-1}{r(r-1)}\sum_{j=1}^r (Y_{ij} - \overline{Y}_{i\cdot})^2 + \frac{\sigma^2_\epsilon}{r}
    \end{array}\!\!\right) 
    \xrightarrow[]{\text{d}} \mathcal{N}\left(\bzero,\tau_1^2\left(\begin{array}{cc}
      1 &  -r^{-1}\\
     -r^{-1}  & r^{-2}\end{array}\right)\right),
\end{align}
as $a\to\infty$, where $\tau_1^2$ is the variance of $\mfrac{1}{r-1}\sum_{j=1}^r (Y_{ij} - \overline{Y}_{i\cdot})^2$, given by
\begin{align}
    \tau_1^2 =\mfrac{1}{(r-1)^2}\,\text{Var}\Big[\sum_{j=1}^r (\epsilon_{ij} - \overline{\epsilon}_{i\cdot})^2\Big].
\end{align}
For the second term, the asymptotic distribution of the sample variance gives,
\begin{align}\label{eq:vec2}
\sqrt{a}\left(\mfrac{1}{a}\sum_{i=1}^a\left(\overline{Y}_{i\cdot} - \overline{Y}_{\cdot\cdot}\right)^2-\sigma^2_u -\mfrac{\sigma^2_\epsilon}{r}\right)
\xrightarrow[]{\text{d}} \mathcal{N}(0,\,\tau_2^2)
\end{align}
as $a\to \infty$, where $\tau_2^2 $ is the variance of $(\overline{Y}_{i\cdot}-\mu)^2$, expressed by, 
\begin{align*}
    \tau_2^2 &= \text{Var}\left[\left(U_i+\overline{\epsilon}_{i\cdot}\right)^2\right].
\end{align*}
Under the one-way model~\eqref{eq:one-way-model} with normality, the variances $\tau_1^2 $ and $\tau_2^2 $ simplify to. 
\begin{align}
\tau_{1}^2 &= \frac{2\sigma^4_\epsilon}{r-1},\\
\tau_2^2 &= 2(\sigma^2_u+r^{-1}{\sigma^2_\epsilon})^2.
\end{align}
Additionally, the two random vectors presented in~\eqref{eq:two-vec} are independent. Combining these results leads to the desired asymptotic distribution for the joint estimator. This completes the proof.

\subsubsection*{Proof of Theorem~\ref{Theorem:a-fixed}}
(a) 
The random variables $\epsilon_{ij}$'s are i.i.d.. Using the consistency of the second sample moment, 
\begin{align}
    \mfrac{1}{r} \sum_{j=1}^r(\epsilon_{ij} - \overline{\epsilon}_{i\cdot})^2 \xrightarrow[]{\text{a.s}} \sigma^2_\epsilon 
\end{align}
as $r\to \infty$ and for any fixed number of units. By summing the above result for all units, we obtain
\begin{align}
     \mfrac{1}{ar} \sum_{i=1}^a\sum_{j=1}^r(\epsilon_{ij} - \overline{\epsilon}_{i\cdot})^2 \xrightarrow[]{\text{a.s}} \sigma^2_\epsilon \, \quad\textrm{ as  } r\rightarrow \infty \, {\textrm{ and }} \forall a \,.
\end{align}
Therefore, $ \widehat{\sigma^2_\epsilon\,} \xrightarrow[]{\text{a.s}} \sigma^2_\epsilon$ as $r\to\infty$.

(b) From the asymptotic normality of the sample variance, we have, 
\begin{align}\label{}
\sqrt{r\,}\left(\mfrac{1}{r}\sum_{j=1}^r\left(\epsilon_{ij} - \overline{\epsilon}_{i\cdot}\right)^2-\sigma^2_\epsilon\right)
\xrightarrow[]{\text{d}} \mathcal{N}(0,\,\tau_\epsilon^2), \;\;\textrm{as } r\to \infty. 
\end{align}
Since $\sum_{j=1}^r(\epsilon_{1j}-\overline{\epsilon}_{1\cdot})^2$, $\sum_{j=1}^r(\epsilon_{2j}-\overline{\epsilon}_{2\cdot})^2,\ldots$ are independent random variables that are normally distributed as $r$ increases, their sum will also be normally distributed. Thus, using the properties of the normal distribution, we get
\begin{align}\label{}
\sqrt{ar}\left(\mfrac{1}{ar}\sum_{i=1}^a\sum_{j=1}^a\left(\epsilon_{ij} - \overline{\epsilon}_{i\cdot}\right)^2-\sigma^2_\epsilon\right)
\xrightarrow[]{\text{d}} \mathcal{N}(0,\,\tau_\epsilon^2), \;\;\textrm{as } r\to \infty. 
\end{align}
which concludes the result.

(c) To prove this part, we start by expressing the maximum likelihood estimate of $\sigma^2_u$ as, 
\begin{align}\label{eq:sigma_U_a_fixed}
    \widehat{\sigma^2_{u}} = \mfrac{1}{a}\sum_{i=1}^a\left(\overline{Y}_{i\cdot} -\overline{Y}_{\cdot\cdot}\right)^2 -\mfrac{1}{r}MS_\epsilon
\end{align}
 We know that the random variable $\sum_{i=1}^a\left(\overline{Y}_{i\cdot} -\overline{Y}_{\cdot\cdot}\right)^2/(\sigma^2_u+r^{-1}\sigma^2_\epsilon)$ follows a chi-squared distribution with $a-1$ degrees of freedom assuming the normal distribution of random effects. Therefore, $a^{-1}\sum_{i=1}^a\left(\overline{Y}_{i\cdot} -\overline{Y}_{\cdot\cdot}\right)^2$ has a gamma distribution with shape parameter ${(a-1)}/{2}$ and scale parameter $2(\sigma^2_u+r^{-1}\sigma^2_\epsilon)/a$.
Moreover, part (b) of this theorem implies that $MS_\epsilon =O_p\left(r^{-{1/2}}\right)$, which allows us to express \eqref{eq:sigma_U_a_fixed} as
\begin{align}
    \widehat{\sigma^2_{u}} = {\textrm{Gamma}}\left(
    \mfrac{a-1}{2},{2a^{-1}\left(\sigma^2_u+r^{-1}\sigma^2_\epsilon\right)} \right)+o_p\left(1\right).
\end{align}
Here, $\textrm{Gamma}(\kappa,\theta)$ denotes the gamma distribution with shape parameter $\kappa$ and scale parameter $\theta$. Since $r^{-1}\sigma^2_\epsilon$ converges to zero as $r\to \infty$, we have that $\widehat{\sigma^2_{u}}$ converges in distribution to a gamma distribution with shape parameter ${(a-1)}/{2}$ and scale parameter $2\sigma^2_u/a$, as $r \to \infty$, i.e.,
\begin{align}
\widehat{\sigma^2_{u}} \xrightarrow[]{\text{d}} \textrm{Gamma}\left(
\mfrac{a-1}{2},{2a^{-1}\sigma^2_u}\right).
\end{align}
 which is the expected convergence result. 

\subsubsection*{Proof of Theorem~\ref{Theorem:positive-sigma-s}}
In the limit case where the sample size $a $ tends to infinity, Theorems~\ref{Theorem:r-fixed} and~\ref{Theorem:ch2-rho} establish that $\widehat{\sigma^2_u}$ converges to $ \sigma^2_u\,$, $\widehat{\sigma^2_\epsilon\,}$ converges to $ \sigma^2_\epsilon$, and $\widehat{\rho}_{_{\scriptsize\mathrm{MLE}}}$ converges to $ \rho $. Furthermore, as $\beta \to 1$, the estimates obtained through the maximum likelihood method and (non-negative) ANOVA method become identical.\smallskip

When the replication number $r$ approaches infinity the estimates take different forms. Using the maximum likelihood method,  $\widehat{\sigma^2_u} \to \sum_{i=1}^a\left(\overline{y}_{i\cdot} -\overline{y}_{\cdot\cdot}\right)^2/a$, $\widehat{\sigma^2_\epsilon\,} \to\sigma^2_\epsilon$, and $\widehat{\rho}_{_{\scriptsize\mathrm{MLE}}} \to \sum_{i=1}^a(\overline{y}_{i\cdot} -\overline{y}_{\cdot\cdot})^2/(a\sigma^2_\epsilon)$.
Under the (non-negative) ANOVA method, the estimates take slightly different forms: $\widehat{\sigma^2_u} \to\sum_{i=1}^a\left(\overline{y}_{i\cdot} -\overline{y}_{\cdot\cdot}\right)^2/(a-1)$, $\widehat{\sigma^2_\epsilon\,} \to\sigma^2_\epsilon$, and $\widehat{\rho}_{_{\scriptsize\mathrm{(N)ANOVA}}}$ approaches $\sum_{i=1}^a\left(\overline{y}_{i\cdot} -\overline{y}_{\cdot\cdot}\right)^2/((a-1)\sigma^2_\epsilon)\,$. In either cases, all estimates remain within the parameter space for their respective parameters.

\end{document}